\documentclass[12pt]{article}
\usepackage[utf8]{inputenc}
\usepackage[english]{babel}
\usepackage[margin=1in]{geometry}
\usepackage{amsmath,amssymb,amsthm}
\usepackage{graphicx}
\usepackage{natbib}
\usepackage{setspace}
\usepackage[colorinlistoftodos]{todonotes}
\usepackage{algorithm}
\usepackage{algpseudocode}
\usepackage{tikz}
\usetikzlibrary{positioning,calc,arrows.meta,shapes}

\usepackage{subcaption}

\usepackage{ mathrsfs }
\usepackage{amsfonts}

\usepackage{bm}

\newcommand{\cm}[1]{\ignorespaces}

\usepackage{booktabs,multirow,tabularx}
\usepackage{graphics}
\usepackage{graphicx}

\usepackage{xcolor}
\definecolor{mypink}{RGB}{219, 48, 122}

\definecolor{mypurple}{RGB}{75,0,130}
\newcommand\yxnew[1]{{\color{magenta} {#1}}} 

	\newtheorem{proposition}{Proposition} 

\newcommand{\cS}{\mathcal{S}}

\newcommand{\cE}{\mathcal{E}}

\newcommand{\bE}{\mathbb{E}}
\newcommand{\bN}{\mathbb{N}}

\newcommand{\Var}{\mathrm{Var}}

\newcommand{\CorT}{\text{CorT}}
\newcommand{\Beta}{\text{Beta}}

\newcommand{\Tree}{\mathrm{Tree}}

\newcommand{\IG}{\mathrm{InvGamma}}

\newcommand{\GEM}{\mathrm{GEM}}
\newcommand{\Bin}{\mathrm{Bin}}
\newcommand{\logit}{\mathrm{logit}}

\newcommand{\br}[1]{\left\{ #1 \right\} }
\newcommand{\mbr}[1]{\left[ #1 \right] }
\newcommand{\sbr}[1]{\left( #1 \right) }

\newcommand\numberthis{\addtocounter{equation}{1}\tag{\theequation}}

\def\PG{\text{PG}}

\def\logit{\text{logit}}

\def\bfI{\mathbf I}
\def\bfX{\mathbf X}

\def\bftheta{\boldsymbol \theta}

\def\bfmu{\boldsymbol\mu}
\def\bftau{\boldsymbol\tau}
\def\bfSigma{\boldsymbol\Sigma}
\def\bfOmega{\boldsymbol\Omega}

\def\bfpsi{\boldsymbol\psi}

\def\bfI{\boldsymbol I}

\def\cS{\mathcal S}
\def\cT{\mathcal T}

\def\iid{\scriptsize \mbox{iid}}
\def\ind{\scriptsize \mbox{ind}}

\def\bfOmega{\boldsymbol\Omega}

\def\diag{\mathrm{diag}}

\def\rT{\mathrm T}
\def\R{\mathbb R}

\newtheorem{definition}{Definition}
\newcommand{\yxreorg}[1]{{#1}}
\renewcommand{\yxnew}[1]{{#1}}

\renewcommand{\CorT}{\text{CorTree}}
\renewcommand{\logit}{\mathrm{logit}}
\def\bfI{\mathbf I}

\usepackage{hyperref}
\hypersetup{colorlinks=true,allcolors=black,
  pdftitle={A tree-based kernel for densities and its applications in clustering DNase-seq profiles},
  pdfauthor={Yuliang Xu, Kaixuan Luo, and Li Ma}}

\newif\ifcodexhighlight
\codexhighlightfalse 
\newcommand{\codexnew}[1]{{%
  \ifcodexhighlight
    \colorlet{codexnewcolor}{magenta}%
  \else
    \colorlet{codexnewcolor}{black}%
  \fi
  \color{codexnewcolor}%
  \hypersetup{allcolors=codexnewcolor}%
  #1%
}}

\title{A tree-based kernel for densities and its applications in clustering DNase-seq profiles}
\author{Yuliang Xu$^{1}$, Kaixuan Luo$^{2}$, and Li Ma$^{3}$%
\thanks{Corresponding author: \href{mailto:li.ma@uchicago.edu}{li.ma@uchicago.edu}.}\\[4mm]
\small $^{1}$Department of Applied Mathematics and Statistics, Johns Hopkins University,\\
\small Baltimore, MD 21218, United States\\[1mm]
\small $^{2}$Department of Human Genetics, University of Chicago,\\
\small Chicago, IL 60637, United States\\[1mm]
\small $^{3}$Department of Statistics \& Data Science Institute, University of Chicago,\\
\small Chicago, IL 60637, United States}
\date{}

\begin{document}
\setstretch{1}
\maketitle
\begin{abstract}
Modeling multiple sampling densities within a hierarchical framework enables borrowing of information across samples. These ``density random effects'' can act as kernels in latent variable models to represent exchangeable subgroups or clusters. A key feature of these kernels is the (functional) covariance they induce, which determines how densities are grouped in mixture models.
Our motivating problem is clustering chromatin accessibility profiles from high-throughput DNase-seq experiments to detect transcription factor (TF) binding. TF binding typically produces footprint profiles with spatial patterns, creating long-range dependency across genomic locations. Existing nonparametric hierarchical models impose restrictive covariance assumptions and cannot accommodate such dependencies, often leading to biologically uninformative clusters.
We propose a nonparametric density kernel that is flexible enough to capture diverse covariance structures and adapts to various spatial patterns of TF footprints. The kernel specifies dyadic tree splitting probabilities via a multivariate logit-normal model with a sparse precision matrix.
Bayesian inference for latent variable models using this kernel is implemented through Gibbs sampling with P\'olya–Gamma augmentation. Extensive simulations show that our kernel substantially improves clustering accuracy. We apply the proposed mixture model to DNase-seq data from the Encyclopedia of DNA Elements project, which results in biologically meaningful clusters corresponding to binding events of two common TFs.
\end{abstract}


\newpage
\setstretch{1.5}

\section{Introduction}\label{sec:intro}

Many modern biomedical applications require modeling multiple unknown groups, each consisting of i.i.d.\ observations drawn from a \yxnew{group-specific} unknown distribution. \yxnew{In particular, we focus on the count data in the form of $\bfX\in \bN_0^{n\times p}$ (see Figure~\ref{fig:heatmap}), where each row is a $p$-dimensional count vector, representing one observation that belongs to an unknown latent group. Such $p$-dimensional random vectors are often assumed to be exchangeable, and $n$ such vectors can be modeled as a mixture of several (group-specific) distributions. Count data are ubiquitous in biological applications using various sequencing technologies: each $p$-dimensional random vector records the count of sequenced fragments assigned to $p$ genomic, taxonomic, or other biological features. For example, in DNase-seq data, the vector records fragment counts across genomic positions around a candidate regulatory site; in microbiome sequencing data, it records counts of DNA or RNA reads assigned to different taxa.} \yxreorg{There are several statistical challenges in clustering such count profiles. The count vectors are usually high-dimensional, overdispersed, and exhibit complex correlation structures driven by biological and spatial factors. Observations of the count vectors often show different distributional patterns across different subgroups, and this forms the basis for differentiating clusters.
}


%



\yxnew{

\noindent\textbf{A brief review of clustering count data.} A variety of methods have been proposed for clustering count vectors. Popular distance-based methods, such as K-means and Partitioning Around Medoids (PAM), cluster observations based on geometric dissimilarity, but do not explicitly account for the complex dependence and overdispersion often present in count data. Model-based approaches provide a more principled alternative by introducing cluster-specific sampling distributions, with either parametric or nonparametric kernels.

Among model-based methods, \cite{holmes2012} adopted a mixture of Dirichlet kernels for each cluster, which accommodates within-cluster overdispersion under a restricted covariance structure. Extending the vector to possibly involve an infinite number of elements, the nested Dirichlet process \citep{rodriguez2008nested} also adopts a mixture model of the Dirichlet kernel, but assumes no heterogeneity within each cluster. \cite{christensen2020bayesian} and \cite{mao2022dirichlet} considered a mixture model with a finite P\'olya tree kernel \citep{ferguson1973bayesian,lavine1992some,dennis1991dt}, which allows within-cluster heterogeneity, but the within-cluster correlation \yxnew{across different elements in the count vector is prespecified by the dyadic tree structure rather than learned adaptively from the data}. More recent works have proposed a mixture of DP mixtures \citep{zhang2025bayesian} with a fixed number of groups, and a covariate-dependent tail-free process \citep{flores2025clustering} to utilize covariate information in clustering.
}

\yxnew{Closely related to our work is the broader literature on hierarchical modeling of exchangeable random distributions. Extending the classical Gaussian hierarchical paradigm, Bayesian nonparametric hierarchical models allow both observation-level and population-level distributions to lie in much richer families. Examples include} \yxreorg{the hierarchical Dirichlet process \citep{teh2006HDP}, hierarchical Pitman-Yor processes \citep{teh2006HPYP,camerlenghi2017bayesian}, more recent hierarchical normalized completely random measures \citep{argiento2020hierarchical}, and hierarchical P\'olya tree \citep{christensen2020bayesian}. While these models have large marginal support for each \yxnew{observation-level distribution},
}they all impose very restrictive constraints on the variation among the \yxnew{observation-level distributions} due to the small number of parameters they devote to characterizing such functional covariance.

\yxnew{\noindent\textbf{DNase-seq data as one motivating example.} DNase-seq is a high-throughput assay for measuring chromatin accessibility.}
\yxreorg{The DNase I enzyme preferentially cuts accessible genomic regions that are not tightly bound by nucleosomes or other proteins, such as transcription factors (TFs). After sequencing and aligning the resulting fragments, one can identify regions with high cleavage intensity, known as DNase I hypersensitive sites, which are often enriched for regulatory elements, including promoters, enhancers, and TF binding sites \citep{encode2012integrated}.}

\yxnew{In our application, the observational units are candidate TF binding sites. Each site is represented by a count profile across genomic positions in a window around the motif match. Within accessible chromatin, TF binding can protect local DNA from DNase I cleavage, producing characteristic \textit{footprints} (Figure~\ref{fig:heatmap}): lower read counts in the motif region and higher counts in the flanking regions, exhibiting a bimodal shape. Clustering such DNase-seq count profiles can therefore help distinguish different patterns of TF binding and improve the identification of bound versus unbound sites. This application also illustrates the statistical challenges of count-data clustering. The count vectors are high-dimensional, overdispersed, and spatially dependent across nearby genomic positions, while substantial heterogeneity may still exist within each cluster among observation-level count vectors. An effective clustering method must therefore incorporate both cross-group and within-group variation while accounting for dependence across the support.}

\yxnew{Several methods have been developed specifically for identifying TF binding sites using DNase-seq data. These include site-centric approaches such as early k-means-based methods \citep{boyle2011high}, covariate-assisted mixture models such as CENTIPEDE \citep{pique2011accurate}, PIQ \citep{sherwood2014discovery}, and segmentation-based methods such as \cite{gusmao2014detection}. While useful in this domain, these methods rely on DNase-seq-specific assumptions or auxiliary information, which makes them less transferable to more general count-data clustering problems, such as microbiome count data.}

\noindent\textbf{Our contribution.} In this paper, we propose a general-purpose mixture model for clustering count vectors \yxnew{with flexible within-cluster covariance, allowing the correlation pattern within each cluster to adapt to the data through a Bayesian prior on the covariance structure.} We develop an efficient sampling algorithm with an R package implementation, and demonstrate its advantages in both simulated and real data. We apply the model to the Encyclopedia of DNA Elements (ENCODE) Consortium \citep{encode2012integrated} DNase-seq data to cluster TF footprints, where it achieves improved accuracy relative to benchmark methods and aligns with the external ChIP-seq reference used for validation. \yxnew{An extension of the proposed method to microbiome data, featuring a phylogenetic tree structure, is applied to the American Gut Project \citep{mcdonald2018american} data in Supplementary Section~\ref{supp_sec:AGP}.}

\noindent\textbf{Organization.} Section \ref{sec:method} introduces our proposed correlated tree model and its theoretical properties. Section \ref{sec:computation} provides details on the sampling algorithm. Section \ref{sec:simulation} provides numerical evidence of the superior performance of our proposed methods based on simulation studies. Section \ref{sec:RDA} gives the biological background of the DNase-seq data and the numerical comparison of our method with the competing methods and the external ChIP-seq reference. We conclude with a discussion on the potential extensions and limitations in Section \ref{sec:discussion}.

\section{A tree-based sampling model for count histograms with flexible covariance}\label{sec:method}

Our proposed sampling model is based on recursive dyadic partitioning of the sample space. At each split, the probability mass is divided between two child nodes, parameterized by the conditional splitting probabilities. We model the conditional splitting probabilities with a prior that accounts for the spatial correlation.

\subsection{The dyadic tree specification of probability distributions}

We begin with a brief introduction to a tree-based parameterization of a probability distribution. This parameterization has been used in both Bayesian \citep{ferguson1973bayesian,lavine1992some} and non-Bayesian \citep{dennis1991dt} literature to model probability distributions. We will later use this parameterization to construct a new model. 

In Figure~\ref{fig:Dyadic_Tree}, we consider a constructive process in which a probability distribution $P$ can be described through sequential splitting of the unit mass along a dyadic partition tree on the sample space $\mathfrak{X}$. Let $\mathcal{T}_0 = \br{\mathfrak{X}}$, $\mathcal{T}_1 = \br{A_0,A_1}$, $\mathcal{T}_2 = \br{A_{00},\dots,A_{11}}$, and so on. Here, we adopt the binary notation of Chapter~3 Section~3.5 of \cite{ghosal2017fundamentals}. Let $l$ be the layer index. $\mathcal{T}_l$ is the $l$-th layer partition consisting of $2^l$ nodes. Each node in $\mathcal{T}_l$ is indexed by a string $\epsilon = \epsilon_1\cdots\epsilon_l$, a length-$l$ sequence of 0s and 1s, indicating whether it is the left (0) or right (1) child in each step of the tree split.
Let $\cE=\br{0,1}$, and let $\cE^l$ be the collection of strings $\epsilon_1\dots\epsilon_l$ of length $l$, which is the index set for $\mathcal{T}_l$, and $\cE^* = \cup_{l=0}^\infty \cE^l$. 
In each split, $A_\epsilon = A_{\epsilon 0}\cup A_{\epsilon 1}$ and $A_{\epsilon 0}\cap A_{\epsilon 1}=\emptyset$, and a probability measure $P$ would satisfy $P(A_\epsilon) = P(A_{\epsilon 0}) + P(A_{\epsilon 1})$.  The probability mass splits according to the conditional probabilities $V_{\epsilon 0} = P(A_{\epsilon 0}|A_{\epsilon})$, and $V_{\epsilon 1} = P(A_{\epsilon 1}|A_{\epsilon}) = 1-V_{\epsilon 0}$. The probability mass on any node $\epsilon = \epsilon_1\dots\epsilon_l\in \cE^l$ in this dyadic tree can be recovered by $P(A_{\epsilon_1\cdots\epsilon_l}) = \prod_{j=1}^l V_{\epsilon_1\dots \epsilon_j}$. 
Figure~\ref{fig:tree_hist} provides the visualization of this dyadic partitioning process in a numerical example for two different distributions supported on $[0,1]$. The top row is one sample drawn from a mixture of two beta distributions with different modes, and the bottom row is another sample drawn from a distribution with one mode.

Figure \ref{fig:tree_hist} illustrates the recursive partitioning process. At each level $l$, the probability distribution $P$ is approximated by a multinomial probability vector of length $2^l$. If a sample $X_1^P,\dots,X_n^P$ is drawn from an unknown distribution $P$, we can partition the sample space $\mathfrak{X}$ in the aforementioned way and count how many $X_i^P$ fall into each set $A_\epsilon$, denoting $n_P(A_\epsilon) = \sum_{i=1}^n I(X^P_i\in A_\epsilon)$. The empirical conditional splitting probability is $\widehat V_{\epsilon 0} = \frac{n_P(A_{\epsilon 0})}{n_P(A_\epsilon)}$ whenever $n_P(A_\epsilon)>0$.

\begin{figure}[htbp]
  \centering
  \begin{subfigure}[b]{0.3\textwidth}
    \centering
    \resizebox{\linewidth}{!}{%

    \begin{tikzpicture}[
  roundnode/.style={circle, draw, minimum size=0.5cm}, 
  plainnode/.style={draw=none, fill=none, minimum size=0mm, text=black}, 
  level distance=2cm, 
  sibling distance=3cm, 
  edge from parent/.style={draw, thick},
  edge from parent path={(\tikzparentnode) -- (\tikzchildnode)},
  level 2/.style={sibling distance=1.5cm} 
]

\node[roundnode] {$\mathfrak{X}$}
  child { node[roundnode] {$A_0$}
    child { node[roundnode] {$A_{00}$}
      edge from parent node[plainnode, left] {$V_{00}$}
    }
    child { node[roundnode] {$A_{01}$}
      edge from parent node[plainnode, right] {$V_{01}$}
    }
    edge from parent node[plainnode, left] {$V_0$}
  }
  child { node[roundnode] {$A_1$}
    child { node[roundnode] {$A_{10}$}
      edge from parent node[plainnode, left] {$V_{10}$}
    }
    child { node[roundnode] {$A_{11}$}
      edge from parent node[plainnode, right] {$V_{11}$}
    }
    edge from parent node[plainnode, right] {$V_1$}
  };

\end{tikzpicture}
    }
    \caption{Dyadic Partitioning Tree}
    \label{fig:Dyadic_Tree}
  \end{subfigure}
  \hfill
  \begin{subfigure}[b]{0.65\textwidth}
    \centering

    \begin{tikzpicture}[
    node distance=1.5cm and 1cm,
    every node/.style={align=center},
    every path/.style={-stealth}
    ]
    
    \node (PThyper) at (0, 0) {$(\pi_\mu,G)$};
    \node (G) [right=of PThyper] {$(\bfmu,\bfSigma)$};
    \node (PTbase) [below=1cm of G] {$\CorT(\bfmu,\bfSigma,\bftau;L)$};
    \node (Q1) [below left=1cm and 1cm of PTbase] {$Q_1$};
    \node (Q2) [below left=1cm and 0cm of PTbase] {$Q_2$};
    \node (Qdots) [below right=1cm and -0.7cm of PTbase] {$\cdots$};
    \node (Qm) [below right=1cm and 1cm of PTbase] {$Q_n$};
    \node (X1) [below=1cmof Q1] {$X^{1}_{[m_1]}$};
    \node (X2) [below=1cmof Q2] {$X^{2}_{[m_2]}$};
    \node (Xdots) [below=1cmof Qdots] {$\cdots$};
    \node (Xm) [below=1cmof Qm] {$X^{n}_{[m_n]}$};
    
    \draw (PThyper) -- (G);
    \draw (G) -- (PTbase);
    \draw (PTbase) -- (Q1);
    \draw (PTbase) -- (Q2);
    \draw (PTbase) -- (Qm);
    \draw (Q1) -- (X1);
    \draw (Q2) -- (X2);
    \draw (Qm) -- (Xm);

    \end{tikzpicture}
    \caption{A tree-based sampling model with hyperpriors on mean and covariance}
    \label{fig:Cor_Tree}
  \end{subfigure}
  \caption{Graphical illustration of CorTree.}
  \label{fig:model}
\end{figure}

\subsection{The hierarchical tree model with sparse precision structure}\label{subsec:cor_tree}

Let $X^i_{[m_i]}=\br{X^i_{1},\dots, X^i_{m_i}}$ be the $i$-th sample with $m_i$ independent draws from an underlying distribution $Q_i$. Within each subgroup, the individual densities $Q_i$ are assumed to independently follow the same underlying sampling distribution, in our case, the correlated tree distribution. In the observed count data, we only have count vectors of $\bfX_i = \br{X_i(B_1),\dots,X_i(B_p)}$ where $B_1,\dots,B_p$ are the histogram partition bins, and $X_i(B_j) = \sum_{k=1}^{m_i}I(X^i_k\in B_j)$. With $n$ independent samples, the data matrix is $\bfX = \sbr{\bfX_1,\dots,\bfX_n}^\rT\in \R^{n\times p}$. In practice, $B_1,\dots,B_p$ are not necessarily the same as the tree leaf nodes, since we use the dyadic partition, and $p$ might not be exactly $2^l$, in which case we usually choose the number of layers $l$ so that $2^l$ is larger than $p$. In this work, to incorporate the spatial structure, we propose a hierarchical tree model with a sparse precision matrix structure. The hierarchical structure is illustrated in Figure \ref{fig:Cor_Tree}, where $Q_i$ is represented by a tree-structured density learner defined below.

Definition \ref{def:tree} formally defines a nonparametric tree model indexed by the transformed splitting probabilities $\bfpsi = \br{\psi_\epsilon,\epsilon \in \cE^*}$. Let $\logit(p):=\log(\frac{p}{1-p})$, and $\logit^{-1}(\psi) = 1/(1+e^{-\psi})$.

\begin{definition}[Dyadic Tree with parameter $\bfpsi$] \label{def:tree}
Given the sample space $\mathfrak{X}$ and a dyadic partition sequence $\{\cT_l\}_{l=0}^\infty$, let $\psi_{\varepsilon} = \logit(V_{\varepsilon 0})$.
Then we say that $X$ follows a Dyadic Tree distribution with parameter $\bfpsi$ and write $X \sim \Tree(\bfpsi)$ if starting from $\epsilon=\emptyset$, $X(A_{\epsilon 0})\,|\,X(A_{\epsilon}),\psi_\epsilon \sim \Bin\br{ X(A_{\epsilon}), \logit^{-1}(\psi_{\epsilon} ) }$ for all $\epsilon$.
\end{definition}

Our main strategy for incorporating spatial correlation is to transform the conditional splitting probability $V_{\epsilon 0}$ to the real line by the logit transform $\psi_{\epsilon} = \logit(V_{\epsilon 0})$ and assign a multivariate normal distribution on the vectorized $\psi_\epsilon$ across different locations $A_\epsilon$ \citep{jara2011class,wang2026tree}. The logit transform in Definition \ref{def:tree} enables us to define a tree model with a covariance structure. Let $I_c \subset \cE^*$ be the index set of all correlated splitting probabilities. For computational convenience, we let the splitting probabilities in layers $l=0,\dots,L$ be correlated. Because $V_{\epsilon 1} = 1-V_{\epsilon 0}$, only including the left splitting probabilities in $I_c$ is sufficient. Hence $|I_c| = \sum_{l=0}^L 2^l = 2^{L+1}-1$.

\begin{definition}[Fixed-layer Correlated Tree]\label{def:cor_tree}
  $\bfpsi$ is said to have a Fixed-layer Correlated Tree distribution if for a fixed integer $L\geq 0$, $\bfpsi_L = \br{\psi_{\epsilon}:\epsilon\in I_c} \in \R^{2^{L+1}-1}$,
    \begin{align*}
        \bfpsi_L &\sim N (\bfmu,\bfSigma), \\
        \psi_{\epsilon} &\overset{\ind}{\sim} N(\mu_\epsilon,\sigma^2_\epsilon), \quad \epsilon \in \cE^*\backslash I_c.\numberthis \label{eq:cor_psi}
    \end{align*}
    Let $\bftau=\br{\mu_\epsilon,\sigma^2_\epsilon}_{\epsilon \in \cE^*\backslash I_c}$ be the collection of parameters in the independent tail splitting variables. Denote $\bfpsi\sim \CorT(\bfmu,\bfSigma,\bftau;L)$ as the correlated tree prior specified in Equation~\eqref{eq:cor_psi}.
\end{definition}
This definition applies to both a tree of finite maximum depth and an infinitely deep tree. We assume that beyond level $L$, the tail splitting variables $\psi_{\epsilon}$ are independent for $\epsilon\in \cE^*\backslash I_c$, and $\mu_{\epsilon}$ and $\sigma_\epsilon^2$ both decay to 0 as the layer goes deeper, making the tail splitting variable $\psi_\epsilon$ shrink to 0 in the deeper tree nodes. We assign hyperpriors $\sigma_\epsilon^2\sim \IG(c,1/l_\epsilon)$ where $l_\epsilon$ is the layer of $\epsilon$, $l_\epsilon = |\epsilon|$, $c$ is an adjustable hyperparameter, with a default of 1. Similarly, $\mu_\epsilon\sim N(0,\sigma_\mu^2)$ where $\sigma_\mu^2$ is allowed to decay to 0 for deeper layers. In practice, when we only have a finite tree, $\sigma_\mu^2$ is set to be a small number.

Such a distribution with finite layers of correlated structure, as in Definition \ref{def:cor_tree}, can be used to model both finite histograms and absolutely continuous distributions with respect to a base measure $\lambda$ such as the Lebesgue measure on $\mathfrak{X}=[0,1]$. In the latter case, it has a large support that includes all absolutely continuous measures with respect to $\lambda$. This allows the models to be applied to continuous sampling distributions as well as finite histograms. While in principle, all real data are discretized at some level (e.g., up to the precision of the measurements or recording), it is helpful to have a sampling distribution supported on density functions that can accommodate long-range dependencies such as those induced by symmetry.

\begin{proposition}\label{prop:ab_cont}
    If $P$ is a random measure with transformed splitting variables $\br{\psi_{\epsilon},\epsilon\in \cE^*}$ that follow Definition \ref{def:cor_tree}, and assume that for the tail splitting variables, $|\mu_{\epsilon}|\leq\delta_l$ and $\sigma_\epsilon^2\leq \gamma_l$ for any $\epsilon \in \cE^l$ where $l>L$, such that $\sum_l (\delta_l + \gamma_l) < \infty$,
    then almost every realization of $P$ is absolutely continuous with respect to $\lambda$. In addition, the total variation support of $P$ consists of all probability measures that are absolutely continuous with respect to $\lambda$.
\end{proposition}

The proof (Supplementary Section~\ref{supp_sec:proof}) of this Proposition directly follows by checking the conditions of Theorems 3.16 and 3.19 in \cite{ghosal2017fundamentals}. The absolute continuity result in Proposition \ref{prop:ab_cont} guarantees that realizations of such random measures can form a density. If the splitting variables divide the mass unevenly (i.e., $\mu_\epsilon$ is away from 0), the resulting realizations may concentrate around some points rather than forming a density. Proposition \ref{prop:ab_cont} also ensures that our prior has full support over the space of probability measures that are absolutely continuous w.r.t. the base measure $\lambda$. Both results benefit from having a tail-free tail in Definition \ref{def:cor_tree}. A tail-free process is defined as $\br{V_0} \perp\!\!\!\perp \br{V_{00},V_{10}} \perp\!\!\!\perp\cdots\perp\!\!\!\perp \br{V_{\epsilon 0}: \epsilon \in \cE^{l}} \perp\!\!\!\perp\cdots$, meaning that the splitting variables across different layers are independent of each other. Our proposed prior in Definition \ref{def:cor_tree} benefits from the good theoretical properties of a tail-free process by having a finite number of layers with correlated splitting variables and an infinite sequence of independent tail splitting variables.

For the mean and covariance of the correlated splitting variables $\bfpsi_L \sim N (\bfmu,\bfSigma)$, 
we let $ \pi_\mu$ be $N(0,\sigma_\mu^2\bfI)$ and assume that $G$ is a prior on the covariance induced by the sparse precision matrix prior. One convenient model that induces such sparsity in the precision is the Graphical Horseshoe (GHS) Prior for which there exist simple computational recipes. (See details in Section \ref{sec:computation}.)

Lastly, for clustering the samples based on their underlying sampling densities, we consider a Dirichlet process mixture (DPM) model with the correlated tree kernel. While one can describe a DPM in several equivalent ways, here we describe it using its stick-breaking representation \citep{sethuraman1994constructive}. In the model, $Z_i$ is a latent cluster membership variable for observation~$i$. 

\begin{definition}[Dirichlet process mixture of correlated trees]\label{def:cluster_tree}
For $i=1,\dots,n$, 
\begin{align*}
    X_i | Z_i=k, \bfpsi_i^{(k)}&\sim \Tree(\bfpsi_i^{(k)}), \\
    \bfpsi_i^{(k)}|Z_i=k &\sim \CorT(\bfmu^{(k)},\bfSigma^{(k)},\bftau^{(k)};L),\quad k=1,2,\dots,\\
    Z_i\,|\,\pi&\sim {\rm MultiBern}(\infty,\pi), \quad\pi \sim \GEM(\alpha), 
    \numberthis \label{eq:cluster_tree}
\end{align*}
where $\GEM(\alpha)$ is the Griffiths-Engen-McCloskey stick-breaking model that induces the Dirichlet process with concentration parameter $\alpha$ \citep{sethuraman1994constructive} and MultiBern($\infty,\pi)$ represents the discrete distribution on the positive integers with probability $\pi_k$ on the value $k$. Specifically, $\pi_k = \beta_k\prod_{b=1}^{k-1}(1-\beta_b), \beta_k\sim \Beta(1,\alpha)$. The superscript \yxnew{$(k)$ in $\bfpsi^{(k)},\bfmu^{(k)},\bfSigma^{(k)},\bftau^{(k)}$ indicates cluster-specific parameters}. We refer to the hierarchical model \eqref{eq:cluster_tree} as the Clustering Correlated Tree.
\end{definition}

Now, we have a correlated tree distribution that (1) incorporates the spatial correlation, (2) is supported on a large class of absolutely continuous distributions, and (3) can be adopted as a distributional kernel in mixture models for clustering. We next turn to computational strategies for posterior sampling under this model. 

\section{Computation}\label{sec:computation}

We implement the Dirichlet process mixture of correlated trees model using a Gibbs sampler, and the R package CorTree is written using RcppArmadillo \citep{RcppArmadillo} and is available on \url{https://github.com/yuliangxu/CorTree}. The posterior sampling of $\bfpsi$ uses the P\'olya-Gamma augmentation \citep{polson2013bayesian}. This allows us to have fully conjugate conditionals on $\bfpsi$. The detailed posterior derivation can be found in the Supplementary Section \ref{supp_sec:derivation}. \yxnew{To approximate the infinite clusters in the stick-breaking process, we truncate to a finite $K$ larger than the true number of clusters \citep{ishwaran2001gibbs}. From a computational perspective, the extra, weakly occupied components provide temporary states to place the uncertain samples during posterior sampling, and the extra components are expected to receive negligible posterior weight after convergence.}

In Section \ref{sec:method}, the prior on the latent precision $\bfOmega$ can be any general sparse precision prior. The sparse precision prior is a family of distributions that has support on the symmetric positive definite matrices and also induces sparsity on the off-diagonal elements. There are also frequentist alternatives for estimating the precision matrix using graphical lasso \citep{friedman2008sparse} and the smoothly clipped absolute deviation penalty \citep{fan2001variable}. Such Bayesian priors include the graphical lasso prior \citep{Wang2012glasso} and the GHS prior \citep{li2019graphical}.
Here, we choose the most recent state-of-the-art sparse precision prior, the Graphical Horseshoe prior, which has been shown in \cite{li2019graphical} to have superior performance in identifying nonzero off-diagonal precision entries.

The GHS prior is placed on the precision matrix $\bfOmega = \bfSigma^{-1} \in \cS_D$, $D=2^{L+1}-1$, where $S_D$ is the space of $D\times D$ positive definite matrices. For $i,j=1,\dots,D$, the GHS prior is defined by assigning horseshoe priors on the off-diagonal elements
\begin{align*}
    p(\omega_{ii})&\propto 1,\quad \omega_{ij:i<j}\sim N(0,\lambda_{ij}^2\tau_\Omega^2),\quad\lambda_{ij}\sim C^+(0,1),\quad \tau_\Omega\sim C^+(0,1),
\end{align*}
where $C^+(0,1)$ is the half-Cauchy distribution. The prior on $\bfOmega$ can be written as 
\[p(\bfOmega,\{\lambda_{ij}\}_{i<j}|\tau_\Omega) \propto \prod_{i<j}N(\omega_{ij}|0,\lambda^2_{ij}\tau_\Omega^2)\prod_{i<j}C^+(\lambda_{ij}|0,1)1_{\bfOmega\in \cS_D}.\]
Algorithm 1 in \cite{li2019graphical} provides the conjugate updates of the full conditionals. We provide a C++ implementation of Algorithm 1 in our package based on the MATLAB implementation from the original paper \citep{li2019graphical}.

\noindent\textbf{Remark.} \codexnew{The original flat-diagonal GHS specification \citep{li2019graphical} is improper; our implementation (\url{https://github.com/yuliangxu/CorTree}) mitigates numerical precision inflation through a covariance-update stopping safeguard without making the prior proper, and comparisons with proper alternatives are reported in Supplementary Section~\ref{supp_sec:proper_priors} and the repository.}

\begin{figure}[ht]
  \centering
  \begin{subfigure}[b]{\textwidth}
    \centering
    \includegraphics[width=\textwidth]{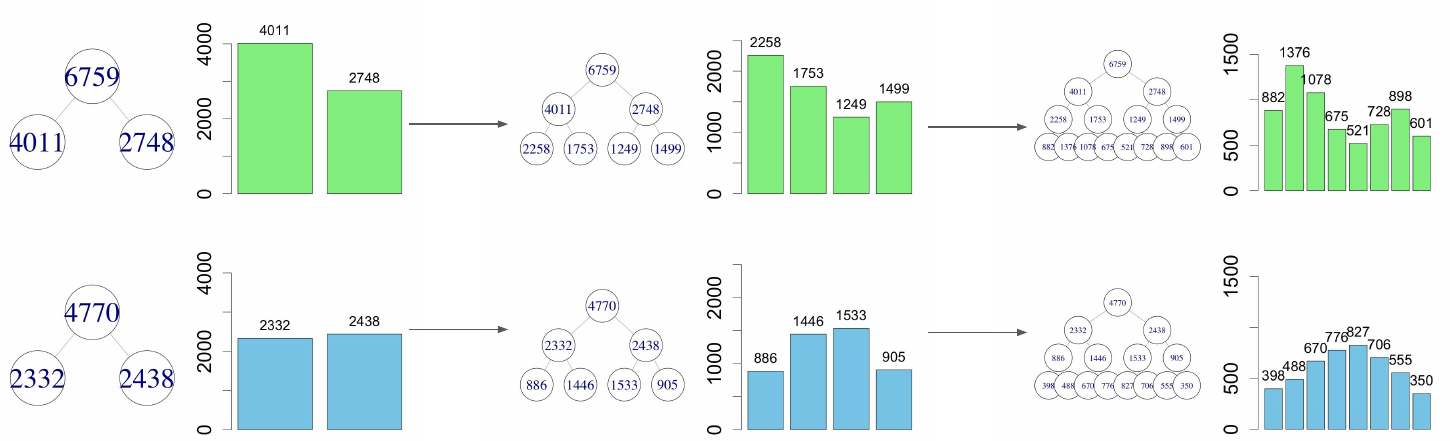}
    \caption{Recursive partition of the sample space via dyadic tree structure for two different densities.}
    \label{fig:tree_hist}
  \end{subfigure}

  \vspace{1em} 

  \begin{subfigure}[b]{\textwidth}
    \centering
    \includegraphics[width=0.7\linewidth]{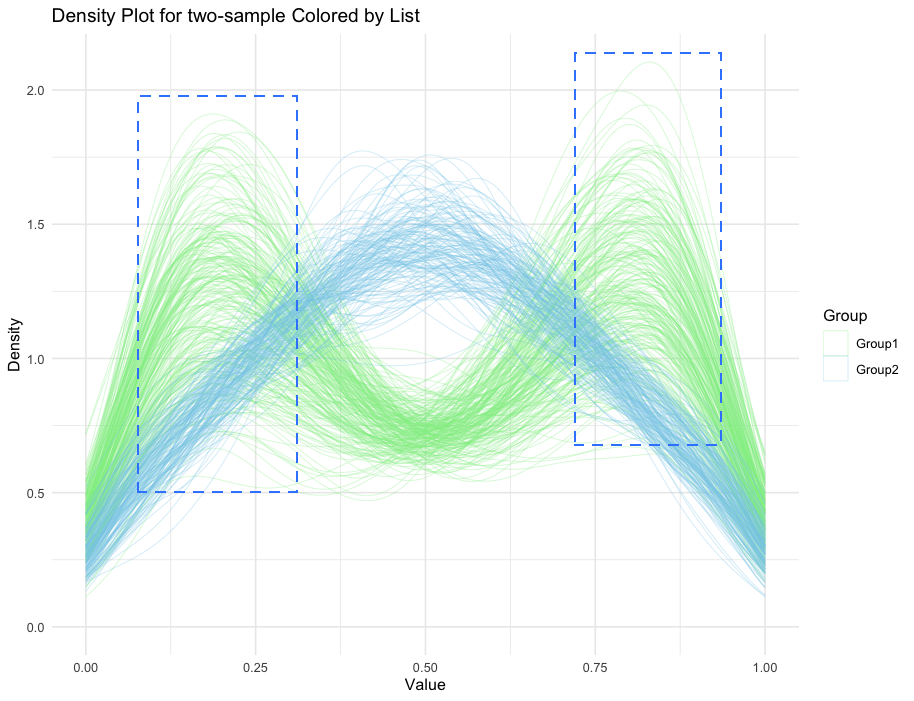}
    \caption{Density plot for the simulated two groups of samples. The green group with two modes (dashed-line boxed) is designed to mimic the transcription factor (TF) ``footprint" pattern.}
    \label{fig:two_sample_density}
  \end{subfigure}

  \caption{Illustration of the simulated data example.}
  \label{fig:both}
\end{figure}

\section{Simulation Study}\label{sec:simulation}

We design a simulation study to compare the clustering performance \yxnew{of CorTree} with the following competing methods: (i) K-means \citep{forgy1965cluster}; (ii) PAM, implemented in R package \texttt{cluster} \citep{PAM}, (iii) independent tree (IndTree), \yxnew{and (iv) Dirichlet Multinomial Mixture (DMM) \citep{holmes2012}.}
\yxnew{The comparison among DMM, IndTree, and CorTree can be seen as an ablation study shown in Figure~\ref{fig:ablation}: DMM models within-cluster heterogeneity through a Dirichlet-multinomial distribution, IndTree models such heterogeneity through independent tree splitting variables, and CorTree further allows the covariance structure among the splitting variables to be adaptively learned from the data.}

We simulate $X_{[m_j]}^j \overset{\iid}{\sim} Q_j$, where $Q_j$ is sampled from a mixture of the following two groups of samples (shown in Figure \ref{fig:two_sample_density}). 
\begin{align*}
    \text{Group 1: }Q_j &= W_j\times\Beta(2,6) + (1-W_j)\times\Beta(6,2), \quad W_j\sim\Beta(10,10); \\
    \text{Group 2: }Q_j &= W_j\times\Beta(1,1) + (1-W_j)\times\Beta(3,3), \quad W_j\sim\Beta(10,10).
\end{align*}
Group 1 (green lines in Figure \ref{fig:two_sample_density}) is a bimodal distribution whose two beta components have modes at $1/6$ and $5/6$ within the support $[0,1]$, mimicking the TF ``footprint" pattern in the DNase-seq data, with a dip in the middle motif region and higher density in the flanking regions around the motif (the two modes as dashed lines boxed in Figure \ref{fig:two_sample_density}). Group 2 (blue lines in Figure \ref{fig:two_sample_density}) is generated as a unimodal distribution. For each density $Q_j$, we sample $X_{[m_j]}^j \overset{\iid}{\sim}Q_j$ with $m_j$ sampled uniformly from integers \yxnew{$10^3$ to $5\times 10^3$}. Each $Q_j$ is a mixture of two beta distributions, with the mixing probability $W_j$ randomly sampled from $\Beta(10,10)$ \yxnew{to represent the within-cluster heterogeneity} of $Q_j$. Each line in Figure \ref{fig:two_sample_density} represents one \yxnew{sampled density} $Q_j$. Some $Q_j$ have a higher left mode than the right mode, others vice versa. We generate a total of $n$ observations $Q_j,j=1,\dots,n$, 60\% from cluster 1 and 40\% from cluster 2. After generating $X_{[m_j]}^j$, we compute the histogram of $X_{[m_j]}^j$ with 1000 bins over $[0,1]$, and use the resulting data as input for clustering algorithms. This gives us $X\in \R^{n\times 1000}$ for each replicated study.

In practice, it is difficult to know {\em a priori} how many true clusters exist in the data, and the competing methods all require a prespecified number of clusters. Hence, we set the cluster number to be 3 for all methods, although there are only 2 true clusters. To quantify the clustering accuracy, we use the Adjusted Rand Index (ARI), defined in \cite{halkidi2002cluster} and implemented in the R package \texttt{mclust} \citep{mclust}. ARI can handle label permutation and provides the corrected-for-chance version of the Rand index. 
ARI maps the clustering similarity to $[-0.5,1]$, with 1 being perfect clustering, 0 indicating agreement expected by chance, and negative values being worse than expected by chance.

The IndTree and CorTree models are implemented using the Gibbs sampler, and the initial value for the membership variable $Z_i$ is set to be the \yxnew{discretized \textit{rowsum} of $X$}. We run 100 MCMC iterations for the burn-in period, and collect the results based on an additional 50 iterations. Both tree models use a six-layer dyadic tree, and the CorTree model uses the splitting variables in layers 0 to 4 to be included in the covariance structure, resulting in a covariance matrix in $\R^{31\times 31}$.

\begin{table}[ht]
\centering
\caption{Clustering results based on 100 replications. Columns report the mean, standard deviation, and median of ARI over 100 replications, the number of times $\mathrm{ARI}=0$ or $\mathrm{ARI}=1$, and the average computation time (in seconds). \yxnew{For each sample, the total count is uniformly drawn from the integers in $[10^3,5\times 10^3]$.} Boldface identifies the highest mean ARI and the largest number of perfect-clustering replications for each sample size.}
\label{tb:Sim_clus}
\resizebox{\columnwidth}{!}{
\begin{tabular}{llrrrrrr}
\hline
Experiment & Method & Mean ARI & S.D.\ ARI & Median ARI & \# of ARI=0 & \# of ARI=1 & Avg.\ time (s) \\
\hline
\multirow{5}{*}{$n=200$} & K-means & 0.30 & 0.04 & 0.30 & 0 & 0 & 0.27 \\
 & PAM & 0.34 & 0.05 & 0.34 & 0 & 0 & 0.06 \\
 & DMM & 0.81 & 0.15 & 0.76 & 0 & 5 & 92.60 \\
 & IndTree & 0.89 & 0.06 & 0.89 & 0 & 0 & 220.71 \\
 & CorTree & \textbf{0.94} & 0.13 & 1.00 & 0 & \textbf{76} & 223.32 \\
\hline
\multirow{5}{*}{$n=400$} & K-means & 0.30 & 0.03 & 0.30 & 0 & 0 & 0.80 \\
 & PAM & 0.34 & 0.04 & 0.34 & 0 & 0 & 0.22 \\
 & DMM & 0.76 & 0.15 & 0.67 & 0 & 3 & 241.62 \\
 & IndTree & 0.91 & 0.04 & 0.91 & 0 & 1 & 441.37 \\
 & CorTree & \textbf{0.96} & 0.10 & 1.00 & 0 & \textbf{83} & 443.05 \\
\hline
\multirow{5}{*}{$n=600$} & K-means & 0.30 & 0.02 & 0.30 & 0 & 0 & 1.39 \\
 & PAM & 0.35 & 0.03 & 0.35 & 0 & 0 & 0.48 \\
 & DMM & 0.76 & 0.15 & 0.66 & 0 & 1 & 349.57 \\
 & IndTree & 0.91 & 0.04 & 0.91 & 0 & 2 & 657.55 \\
 & CorTree & \textbf{0.96} & 0.13 & 1.00 & 1 & \textbf{86} & 662.49 \\
\hline
\end{tabular}
}
\end{table}

Based on the results in Table \ref{tb:Sim_clus}, among all competing methods, CorTree achieves the highest mean and median ARI. DMM, IndTree, and CorTree, in some instances, can achieve perfect clustering (ARI=1). \yxnew{The incrementally flexible design in Figure~\ref{fig:ablation} from DMM, IndTree, to CorTree, leads to increasingly better performance.}


\section{Clustering of TF footprints using DNase-seq data from ENCODE}\label{sec:RDA}

\yxnew{As discussed in Section~\ref{sec:intro}}, identifying TF binding sites is critical to understanding gene regulation. We apply \yxnew{CorTree} to DNase-seq data from the ENCODE Consortium \citep{encode2012integrated}, with two key TFs, REST (RE1-silencing TF) and NRF1 (Nuclear respiratory factor 1), in the K562 human erythroleukemia cell line. NRF1 maintains cellular energy metabolism and mitochondrial function. REST is a repressor of neuronal gene expression in non-neuronal cells and modulates neurodevelopment.

\yxnew{Since DNase-seq data do not contain TF binding specificity information, to define the candidate binding sites for our TFs of interest in our analysis, we scan the genome for TF motif matches.} We download position weight matrices (PWMs) from the JASPAR database \citep{rauluseviciute2024jaspar}, scan the human genome for motif matches with REST and NRF1 motifs using FIMO software \citep{grant2011fimo}, select motif matches with FIMO p-value $<$ 1e-5 and PWM scores $\geq$ 13, and take 100-bp flanking regions on both sides of the motif matches as the candidate regions for our analysis \yxnew{(Supplementary Section~\ref{supp_sec:RDA} presents an analysis without the PWM score preprocessing)}. We then combine the DNase counts on both strands, and restrict to candidate \yxnew{sites} with DNase counts $\geq$ 50 for both REST and NRF1 data. \yxnew{After preprocessing, $\bfX_{\text{REST}}\in \bN_0^{4551\times 220}$, with the rowsum ranging from 50 to 9399; $\bfX_{\text{NRF1}}\in \bN_0^{8148\times 211}$, with the rowsum ranging from 50 to 22576.}
\yxnew{To validate the clustering results, we use ChIP-seq data, which have been widely used as a proxy for TF binding truth \citep{park2009chip}. Unlike DNase-seq, ChIP-seq uses an antibody to target a specific TF and therefore provides TF-specific binding measurements. From the ENCODE portal, we obtain normalized ChIP-seq read counts and assign binary labels at the motif match locations that overlap with ChIP-seq peaks. Because ChIP-seq is based on a different experimental technology, we use it only as an independent external reference for validation and method comparison.}

We also include a state-of-the-art method tailored for DNase-seq data called CENTIPEDE \citep{pique2011accurate}, which uses the continuous PWM score as a site-specific covariate. CENTIPEDE uses a simple mixture of two multinomial components in \eqref{eq:CENTIPEDE}, which clusters sites into 0-1 groups and requires additional site-specific information $Y_i$.
\begin{equation} \label{eq:CENTIPEDE}
    \bfX_i \in \bN_0^p, \bfX_i\sim \pi_i\text{Multinomial}(n_i,\bftheta_1) + (1-\pi_i)\text{Multinomial}(n_i,\bftheta_0).\quad \text{logit}(\pi_i) = Y_i^\rT\beta.
\end{equation}
$Y_i$ is a 2-dimensional vector, $(1,\text{PWM\_score}_i)$, to include the intercept term. The unbound sites are simply assumed to be uniformly distributed in CENTIPEDE, which may fail to capture the spatial bias induced by the DNase I enzyme.

Because of the stick-breaking prior for the mixing weights in \eqref{eq:cluster_tree}, we need to set a large number of clusters for CorTree to allow the posterior to collapse to fewer clusters. We set 5 clusters for CorTree \yxnew{and IndTree} to learn potential subgroup structures, 2 clusters for k-means, PAM, \yxnew{DMM}, and CENTIPEDE. Only clusters with more than 10 observations are reported. \yxnew{For CorTree and IndTree,} \yxreorg{the tree structure is set to 9 layers to have at least one leaf node to represent a single column in $\bfX$. This dyadic partition leads to many 0s in deeper layers of the tree.} \yxnew{We set layers 0 to 3 to be correlated for REST data ($\Sigma\in \R^{15\times 15}$), layers 0 to 4 to be correlated for NRF1 data ($\Sigma\in \R^{31\times 31}$), collecting 50 MCMC iterations after 100 burn-in iterations (see traceplots in Figure~\ref{fig:mixing}). The tuning parameters include the number of correlated layers, the hyperparameter $\alpha_0$ in the prior $\sigma_{k,I}^2\sim \IG(\alpha_0,\beta_{0,I})$, and the hyperparameter $\sigma_\mu^2$ in the hyper-prior $N(0,\sigma_\mu^2)$ for $\mu_{k,C},\mu_{k,I}$. Here $I,C$ are the index sets for independent and correlated splitting variables, respectively. Supplementary Section \ref{supp_sec:RDA} provides sensitivity analysis for CorTree under varying initial values and hyperparameters.}


\yxnew{As a nonparametric clustering method with flexible covariance structure, the posterior of CorTree may have many modes, especially when there are thousands of samples: many distinct cluster configurations can provide similarly good explanations of the data. The choice of the initial $Z$ must be scientifically meaningful to guide meaningful clustering. For the TF binding detection task, the most informative choice is the PWM score, which indicates TF binding specificity, and is also used by CENTIPEDE as a predictor. Other informative candidates include the total count of each observation (\textit{rowsum} of the count data), since higher total counts suggest higher chromatin accessibility of the region; and the distance from the motif matches to the nearest transcription start site (TSS; \textit{tss-dist}), which measures the distance from the candidate TF binding sites to the nearest TSS. We show the performance of CorTree and IndTree under the three different initial values: \textit{pwm}, \textit{rowsum}, and \textit{tss-dist} in terms of ARI (Table~\ref{tab:RDA_ARI}), and provide visualizations of results with \textit{pwm} initial value as the final result, with other initial value results in Supplementary~\ref{supp_sec:RDA}.}

\begin{figure}[ht]
  \centering
  \includegraphics[width=\linewidth]{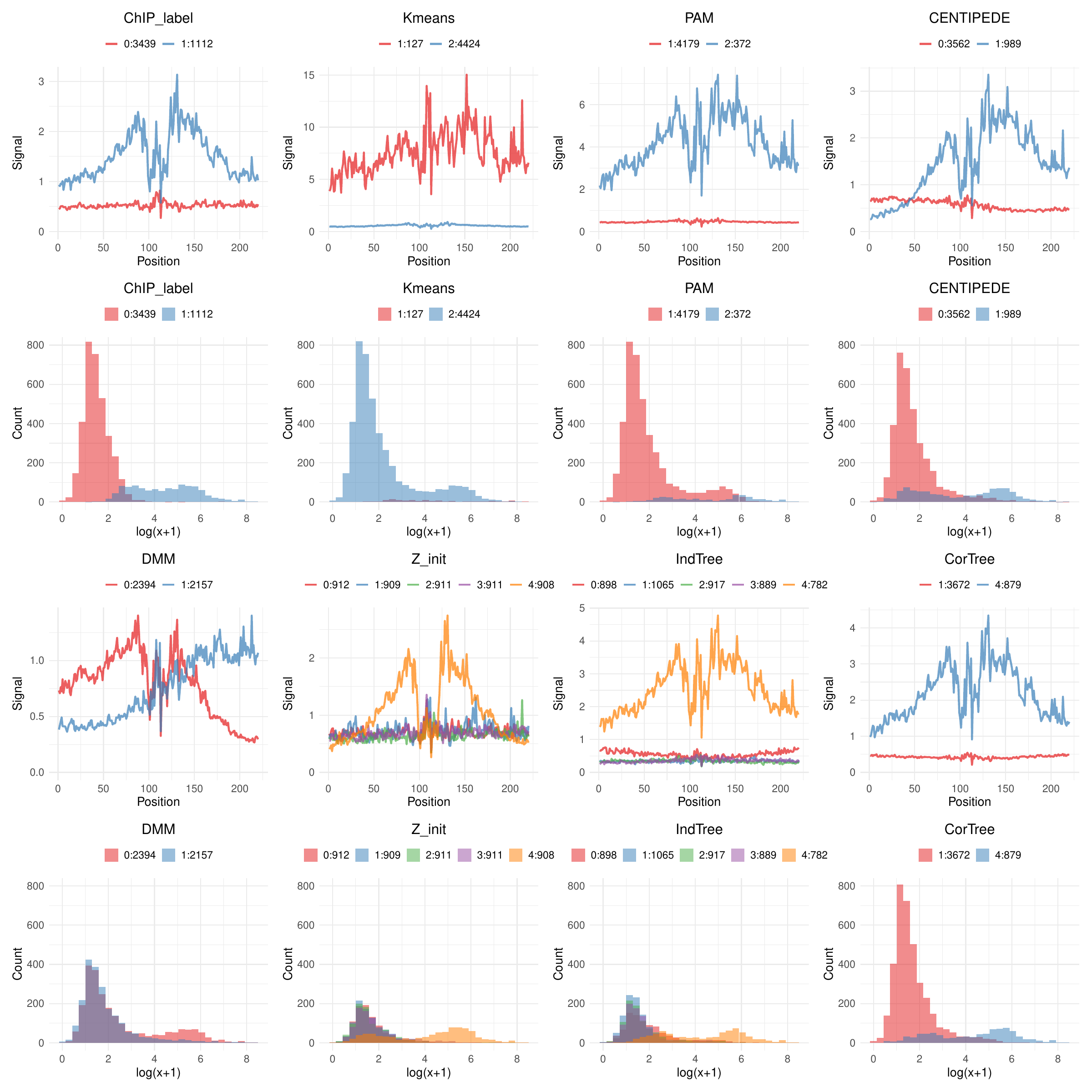}
  \caption{REST data clustering result. K-means, PAM, CENTIPEDE, and DMM are set to have 2 clusters. CorTree uses 5 clusters but converges to 2 clusters after burn-in. The first column at the top is the reference binary result from ChIP data. Rows 1 and 3 are the cluster mean plots, and Rows 2 and 4 are the histograms of the log-transformed ChIP count data. The legend is formatted as \textit{subgroup label:subgroup size}.}
  \label{fig:RDA_REST_2clus}
\end{figure}

\noindent\textbf{Analysis results.}
Figure \ref{fig:RDA_REST_2clus} provides the binary clustering performance. The cluster-mean plots help to check the spatial pattern across genomic locations for each subgroup. They show a clear bimodal pattern for \yxnew{the subgroup} of potential binding sites, and another subgroup with a flat pattern over the genomic locations, potentially the unbound sites. When restricted to only 2 clusters, k-means and PAM both find a small ($<$10\% of the total subsample) subgroup with an overall high count, and put the rest in one big subgroup. CENTIPEDE can find a subgroup with an asymmetric bimodal pattern, which could also be a potential binding group.
The histograms in Figure \ref{fig:RDA_REST_2clus} are the ChIP count data stratified by clusters. They provide a 1-dimensional summary of how likely a candidate site could be a binding site. The binding subgroups identified by K-means and PAM are nearly indistinguishable in the ChIP count distribution. Both CENTIPEDE and CorTree can find a subgroup of potential binding sites that also have high ChIP counts, and the binding subgroups (blue) are more distinguished from the unbound subgroups (red). However, if we compare the left tail of the blue groups for CENTIPEDE and CorTree, CENTIPEDE tends to classify more sites with slightly lower ChIP counts into the bound group.

\begin{figure}[ht]
  \centering
  \includegraphics[width=\linewidth]{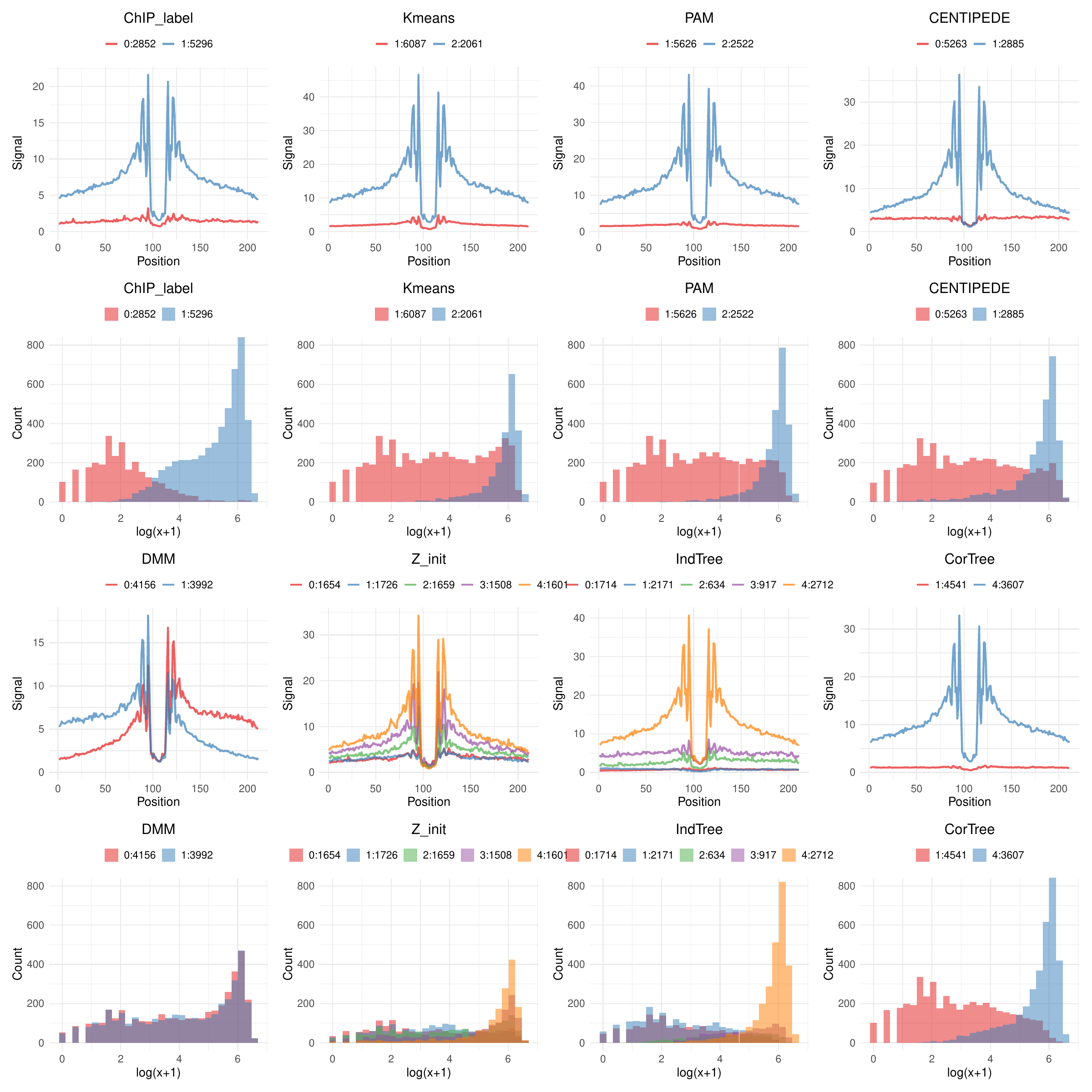}
  \caption{NRF1 data clustering result. K-means, PAM, CENTIPEDE, and DMM are set to have 2 clusters. CorTree uses 5 clusters but converges to 2 clusters after burn-in. The first column at the top is the reference binary result from ChIP data. Rows 1 and 3 are the cluster mean plots, and Rows 2 and 4 are the histograms of the log-transformed ChIP count data. The legend is formatted as \textit{subgroup label:subgroup size}.}
  \label{fig:RDA_NRF1}
\end{figure}

\yxreorg{Figure \ref{fig:RDA_NRF1} shows results for NRF1 data, whose pattern is quite different from the REST pattern. The data became more unbalanced after preprocessing, with more samples in the bound group than the unbound group. The cluster-mean plots in Figure \ref{fig:RDA_NRF1} show that all methods provide similar cluster mean patterns for the bound and unbound groups. Considering both the cluster patterns and group sizes (histograms in Figure \ref{fig:RDA_NRF1}), only CorTree identifies a representative bound-site group with a size comparable to the binary ChIP label (3607 versus 5296).}

\yxnew{Table~\ref{tab:RDA_ARI} provides ARI checks of each method with the reference ChIP label. CorTree outperforms CENTIPEDE on all three initial values in the NRF1 task, and outperforms CENTIPEDE in the REST task with the PWM initial value. Recall that CENTIPEDE also uses PWM scores as the predictor here \yxnew{(Section~\ref{supp_sec:RDA} provides CENTIPEDE results using other predictors like \textit{tss-dist})}. Table~\ref{tab:RDA_ARI} also shows that the converged posteriors of CorTree are already reasonably far from the initial values (last row in Table~\ref{tab:RDA_ARI}) and closer to the external ChIP-seq reference labels.}

\begin{table}[ht]
\centering
\caption{ARI compared with the binary ChIP label, by method and initialization. Boldfaced ARI values are greater than CENTIPEDE's ARI for the same dataset. The CorTree and IndTree blocks report clustering results, respectively, for varying initial values. The final block \textit{Initial values for $Z$} refers to ARI of the initial values compared to the ChIP label.}
\label{tab:RDA_ARI}
\resizebox{\columnwidth}{!}{
\begin{tabular}{lcccccccc}
\hline
 & \multicolumn{4}{c}{\textbf{REST}} & \multicolumn{4}{c}{\textbf{NRF1}} \\
\hline
\textbf{Methods} & K-means & PAM & CENTIPEDE & DMM & K-means & PAM & CENTIPEDE & DMM \\
ARI & 0.069 & 0.217 & 0.336 & 0.012& 0.005 & 0.061 & 0.089 & 0 \\
\hline
 &  & \multicolumn{3}{c}{CorTree} &  & \multicolumn{3}{c}{CorTree} \\
Init &  & rowsum & pwm & tss-dist &  & rowsum & pwm & tss-dist \\
ARI &  & 0.319 & \textbf{0.418} & 0.244 &  & \textbf{0.199} & \textbf{0.203} & \textbf{0.240} \\
\hline
 &  & \multicolumn{3}{c}{IndTree} &  & \multicolumn{3}{c}{IndTree} \\
Init &  & rowsum & pwm & tss-dist &  & rowsum & pwm & tss-dist \\
ARI &  & 0.054 & 0.075 & 0.043 &  & \textbf{0.115} & \textbf{0.157} & \textbf{0.176} \\
\hline
 &  & \multicolumn{3}{c}{Initial values for $Z$} &  & \multicolumn{3}{c}{Initial values for $Z$} \\
Init &  & rowsum & pwm & tss-dist &  & rowsum & pwm & tss-dist \\
ARI &  & 0.058 & 0.087 & 0.001 &  & \textbf{0.118} & 0.038 & 0.031 \\
\hline
\end{tabular}
}
\end{table}

\section{Discussion}\label{sec:discussion}

We introduced CorTree, a hierarchical tree model enriched with sparse precision structure in the splitting variables, thereby accounting for long-range dependencies in multi-group count data. CorTree has large prior support in theory, and the precision matrix is assigned the graphical horseshoe prior to adaptively learn the second moment structure in the MCMC algorithm.  Extensive simulations demonstrate that CorTree substantially outperforms classical clustering methods such as k-means, PAM, \yxnew{DMM}, and the hierarchical tree with independent structure. Combining flexible tree-based density estimation with Bayesian clustering, CorTree provides a general solution to finding latent subgroups in high-dimensional count data. CorTree can be extended to other settings, such as clustering microbiome count profiles. \yxnew{We leverage the phylogenetic tree to incorporate contextual knowledge of the taxa and apply CorTree to the American Gut Project data in Supplementary Section~\ref{supp_sec:AGP}}.  There are two main challenges in extending this method to other domains: a domain-informative initial value for the membership variable; and a relatively good ground truth to evaluate the clustering results. The latter is a common challenge in unsupervised learning tasks such as clustering and requires domain experts to extract meaningful information from the identified subgroups.

\section*{Acknowledgements}

Most of this research was completed when YX and LM were at Duke University and when YX was at the University of Chicago. We have used ChatGPT to check for grammar mistakes and carry out sentence-level edits. \vspace*{-8pt}

\section*{Supplementary Materials}
Supplementary material is available at Biometrics online.

Web Appendices, Tables, and Figures referenced in Sections \ref{sec:intro}--\ref{sec:discussion}, along with data and code are available with this paper at the Biometrics website on Oxford Academic.
The code for the CorTree R package, simulation scripts, and real-data analysis workflows are available as a zip file in the supplementary materials, as well as at \url{https://github.com/yuliangxu/CorTree}. The repository includes the R/Rcpp implementation of the CorTree and PhyloTree samplers, helper functions for model fitting and visualization, and scripts used to reproduce the simulation, DNase, and microbiome analyses.

\section*{Funding}

This research is partly supported by NIGMS grant R01-GM135440.

\section*{Conflicts of Interest}

None declared.

\section*{Data Availability}

PWMs for RE1-silencing transcription factor (REST; MA0138.3) and Nuclear Respiratory Factor 1 (NRF1; MA0506.1) were downloaded from the JASPAR 2024 database \citep{rauluseviciute2024jaspar}. Deoxyribonuclease I hypersensitive-site sequencing (DNase-seq) and chromatin immunoprecipitation sequencing (ChIP-seq) data for REST and NRF1 in the K562 cell line were downloaded from the ENCODE portal \citep{sloan2016encode} (\url{https://www.encodeproject.org/}) with the following identifiers: ENCSR000EMT, ENCSR137ZMQ, and ENCSR837EYC. Data were processed using the Transcription factor Occupancy Profiler (TOP) R package \citep{luo2022profiling}, with more details available at: \url{https://kevinlkx.github.io/footprint_clustering/index.html}.

\bibliographystyle{asa}
\bibliography{ref}

\clearpage
\setstretch{1}
\section*{Supplementary Materials}

\setcounter{figure}{0}
\setcounter{table}{0}
\setcounter{section}{0}
\renewcommand{\thefigure}{S\arabic{figure}}
\renewcommand{\thetable}{S\arabic{table}}
\renewcommand{\thesection}{S\arabic{section}}
\renewcommand{\theHfigure}{S\arabic{figure}}
\renewcommand{\theHtable}{S\arabic{table}}
\renewcommand{\theHsection}{S\arabic{section}}

\section{Proof of Proposition \ref{prop:ab_cont}}\label{supp_sec:proof}

\begin{proof}
We give the argument for the regular dyadic partition of $[0,1]$, with $\lambda$ the uniform (Lebesgue) probability measure, so that $\lambda(A_\epsilon)=2^{-|\epsilon|}$. The same argument applies to a base probability measure with a generating partition that divides its mass equally at every split. We verify finite-dimensional full support and a uniform second-moment bound, and then establish absolute continuity and total variation support directly; see also the discussion of these properties in \cite{ghosal2017fundamentals}.

Let $h(x)=(1+e^{-x})^{-1}$. For a path $\epsilon=\epsilon_1\cdots\epsilon_K$, write $\epsilon_{1:l}=\epsilon_1\cdots\epsilon_l$, with $\epsilon_{1:0}=\emptyset$, and define
\begin{align*}
    U_{\epsilon,l}=V_{\epsilon_{1:l}}
    &=h\sbr{s_l\psi_{\epsilon_{1:l-1}}}, \qquad
    s_l=\begin{cases}1,&\epsilon_l=0,\\-1,&\epsilon_l=1.\end{cases}
\end{align*}
Thus $P(A_\epsilon)=\prod_{l=1}^K U_{\epsilon,l}$. By Definition~\ref{def:cor_tree}, the correlated parent-node depths are $0,\dots,L$, corresponding to path depths $1,\dots,L+1$. The remaining splitting variables are independent of one another and of this finite correlated block.

The correlated Gaussian covariance is positive definite, and the tail variances are positive under the model. Hence every finite vector of transformed splitting variables has full support on its Euclidean space. Applying the logistic map coordinatewise gives full support on the corresponding unit cube. In particular, the vector $\br{V_{\eta 0}:\eta\in\cE^K}$ has support $[0,1]^{2^K}$.

Consider a tail parent node $\eta$ of depth $j=|\eta|>L$ and either branch probability $U=h(s\psi_\eta)$, where $s\in\br{-1,1}$. Write $s\psi_\eta=m+Z$, with $m=s\mu_\eta$ and $Z$ centered normal. Since $h$ is Lipschitz with constant $1/4$,
\begin{align*}
    \Var(U)&\leq \bE\mbr{(h(m+Z)-h(m))^2}
    \leq \frac{1}{16}\bE Z^2
    \leq \frac{\gamma_j}{16}.
\end{align*}
Moreover, symmetry of $Z$ and the identity $h(z)+h(-z)=1$ imply $\bE h(Z)=1/2$. Therefore,
\begin{align*}
    \left|\bE U-\frac12\right|
    &=\left|\bE\sbr{h(m+Z)-h(Z)}\right|\\
    &\leq \bE\left|h(m+Z)-h(Z)\right|
    \leq \frac{|m|}{4}
    \leq \frac{\delta_j}{4}.
\end{align*}
Using the looser bounds $\Var(U)\leq\gamma_j$ and $|\bE U-1/2|\leq\delta_j$, we obtain
\begin{align*}
    4\bE(U^2)
    &=4\Var(U)+4(\bE U)^2\\
    &\leq 4\gamma_j+4(1/2+\delta_j)^2
    =1+a_j,\\
    a_j&=4\gamma_j+4\delta_j+4\delta_j^2.
\end{align*}
The assumption $\sum_j(\delta_j+\gamma_j)<\infty$ also implies $\sum_j\delta_j^2<\infty$, and hence $\sum_j a_j<\infty$.

For $K>L+1$, independence of the tail from the finite correlated block gives
\begin{align*}
    \frac{\bE\sbr{P(A_{\epsilon_1\cdots\epsilon_K})^2}}
         {\lambda(A_{\epsilon_1\cdots\epsilon_K})^2}
    &=\bE\sbr{\prod_{l=1}^{L+1}(2U_{\epsilon,l})^2}
      \prod_{l=L+2}^K 4\bE(U_{\epsilon,l}^2)\\
    &\leq 4^{L+1}\prod_{j=L+1}^{K-1}(1+a_j)\\
    &\leq 4^{L+1}\exp\sbr{\sum_{j>L}a_j}<\infty.
\end{align*}
For $K\leq L+1$, the same ratio is at most $4^K\leq4^{L+1}$. Consequently, this bound is uniform over $K$ and all paths $\epsilon\in\cE^K$.

To establish absolute continuity, define the partition density
\[
    f_K(x)=\sum_{\epsilon\in\cE^K}
    \frac{P(A_\epsilon)}{\lambda(A_\epsilon)}\,\mathbf{1}_{A_\epsilon}(x).
\]
For each realization of $P$, these densities form a nonnegative martingale with respect to the dyadic partitions under $\lambda$. The uniform bound above gives $\sup_K\bE\|f_K\|_{L^2(\lambda)}^2<\infty$. Since $\|f_K\|_{L^2(\lambda)}^2$ is nondecreasing in $K$, it follows that $\sup_K\|f_K\|_{L^2(\lambda)}^2<\infty$ almost surely. Thus $f_K$ converges in $L^2(\lambda)$, and hence in $L^1(\lambda)$, to a density $f$ with the same probabilities as $P$ on every dyadic cell. These cells generate the Borel sets, so $P$ has density $f$ almost surely.

For total variation support, let $g$ be any $\lambda$-density and let $g_K$ be its dyadic partition averages. Then $g_K\to g$ in $L^1(\lambda)$. For $K\geq L+1$, set $B_K=\prod_{j=K}^\infty(1+a_j)$, so $B_K\to1$. Conditional on the splitting variables at parent depths below $K$, the same second-moment bound on each normalized subtree below a depth-$K$ cell gives
\[
    \bE\sbr{\|f-f_K\|_{L^1(\lambda)}\,\big|\,\mathcal F_K}
    \leq \sqrt{B_K-1},
\]
where $\mathcal F_K$ is generated by those finite-prefix splitting variables. Indeed, each normalized subtree density has integral one and expected squared $L^2$ distance from the constant density at most $B_K-1$; summing the resulting $L^1$ bounds with weights $P(A_\epsilon)$ proves the display. For any $\varepsilon>0$, choose $K$ large enough that $\|g-g_K\|_{L^1(\lambda)}<\varepsilon/3$ and $\sqrt{B_K-1}<\varepsilon/6$. Finite-dimensional full support ensures that $\|f_K-g_K\|_{L^1(\lambda)}<\varepsilon/3$ has positive probability. Conditional Markov's inequality gives probability greater than $1/2$ that $\|f-f_K\|_{L^1(\lambda)}<\varepsilon/3$, uniformly over the finite prefix. Their intersection therefore has positive probability, and on it $\|f-g\|_{L^1(\lambda)}<\varepsilon$. This proves full total variation support over the $\lambda$-densities.
\end{proof}

\section{Posterior derivation for Gibbs sampling algorithm}\label{supp_sec:derivation}

\subsection{Posterior of $\psi_i$ and its hyperparameters}

We vectorize the splitting variables from top to bottom and from left to right, as implemented in the \texttt{vectorize\_tree} C++ function in our R package, and denote the resulting vector by $\bfpsi_i$. Let $C=I_c=\bigcup_{j=0}^L\cE^j$ be the set of correlated parent nodes, with $|C|=2^{L+1}-1$, and let $I$ contain the remaining internal nodes of the finite tree used for computation. Thus $\bfpsi_i=(\bfpsi_{i,C}^{\rT},\bfpsi_{i,I}^{\rT})^{\rT}$. Given $Z_i=k$, the correlated block has normal prior with mean $\bfmu_{k,C}$ and precision $\Lambda_{k,C}$. The independent coordinates have normal priors with means $\mu_{k,\epsilon}$ and variances $\sigma_{k,\epsilon}^2$, for $\epsilon\in I$. The GHS prior on $\Lambda_{k,C}$ is updated as described in Section~\ref{sec:computation}.

We use P\'olya--Gamma augmentation \citep{polson2013bayesian} to update $\bfpsi_i$. For a parent node $A_\epsilon$ and its left child $A_{\epsilon 0}$, the transformed splitting variable is $\psi_{i,\epsilon}$, following Definition~\ref{def:tree}. With $n_i(A_\epsilon)$ the count in the parent node, draw
\[
    \omega_i(A_\epsilon)\mid\dots
    \sim\PG\sbr{n_i(A_\epsilon),\psi_{i,\epsilon}}.
\]
The posterior of the correlated block, conditional on $Z_i=k$, is
\begin{align*}
    \bfpsi_{i,C}\mid\dots
    &\sim N\sbr{Q_{i,k,C}^{-1}
        (\Lambda_{k,C}\bfmu_{k,C}+\kappa_{i,C}),\,
        Q_{i,k,C}^{-1}},\\
    Q_{i,k,C}&=\Lambda_{k,C}+\Omega_{i,C},\\
    \Omega_{i,C}&=\diag\sbr{(\omega_i(A_\epsilon))_{\epsilon\in C}},\\
    \kappa_{i,C}&=\sbr{n_i(A_{\epsilon 0})-\frac{n_i(A_\epsilon)}{2}}_{\epsilon\in C}.
\end{align*}
The precision $\Lambda_{k,C}$ may be dense. We use a Cholesky decomposition of $Q_{i,k,C}$ to evaluate the Gaussian update. The independent-block posterior has the same form, replacing $\Lambda_{k,C}$ by $\diag\sbr{(\sigma_{k,\epsilon}^{-2})_{\epsilon\in I}}$, $\bfmu_{k,C}$ by $\bfmu_{k,I}$, and $\Omega_{i,C}$ and $\kappa_{i,C}$ by their counterparts indexed by $I$. We assign the hyperprior $N(0,\sigma_\mu^2)$ to each coordinate of $\bfmu_{k,C}$ and $\bfmu_{k,I}$.

For each independent node $\epsilon\in I$, with parent depth $|\epsilon|>L$, the variance has inverse-gamma prior $\IG(\alpha_0,\beta_{0,\epsilon})$, where $\beta_{0,\epsilon}=1/|\epsilon|$. Its full conditional is
\[
    \sigma_{k,\epsilon}^2\mid\dots
    \sim\IG\sbr{\alpha_0+\frac{n_k}{2},\,
        \beta_{0,\epsilon}+\frac{S_{k,\epsilon}}{2}},
    \qquad
    S_{k,\epsilon}=\sum_{i:Z_i=k}(\psi_{i,\epsilon}-\mu_{k,\epsilon})^2,
\]
where $n_k$ is the number of observations assigned to cluster $k$.

\begin{figure}
    \centering
    \includegraphics[width=0.7\linewidth]{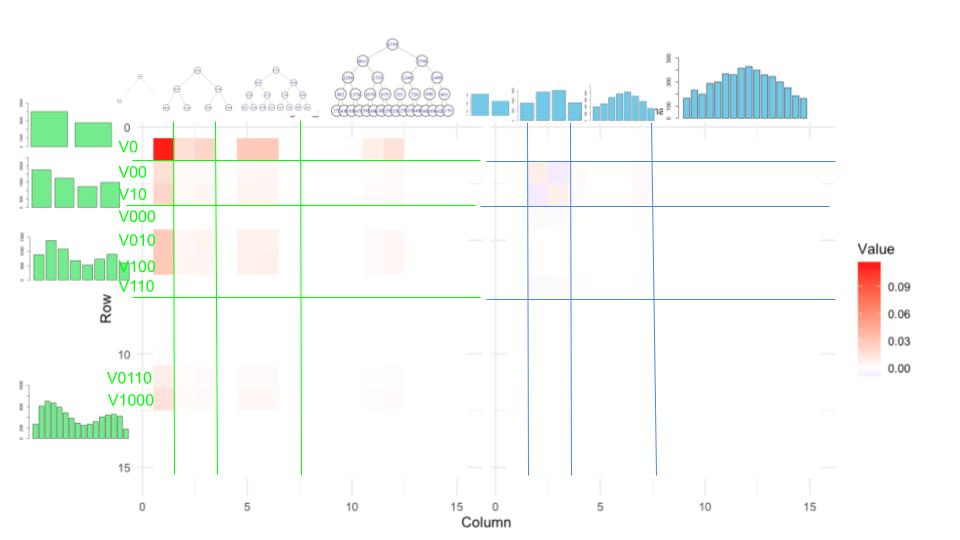}
    \caption{Illustration of the estimated $\Sigma_k$ for the two groups shown in Figure~\ref{fig:two_sample_density}.}
    \label{fig:coarse_to_fine}
\end{figure}

\textbf{Figure~\ref{fig:coarse_to_fine}.} We illustrate the estimated cluster-level covariance matrices $\Sigma_k$ for the two groups in Figure~\ref{fig:two_sample_density}. The sparsity-inducing prior is placed on the precision matrix $\Sigma_k^{-1}$. For Gaussian splitting variables, a zero off-diagonal precision entry represents conditional independence of the corresponding transformed splitting variables given the remaining coordinates. This does not require their marginal covariance to be zero: indirect dependence can produce a dense covariance matrix even when the precision matrix is sparse. The coordinates of $\bfpsi_i$ are ordered from the root to finer tree levels, and the covariance patterns in Figure~\ref{fig:coarse_to_fine} illustrate the different marginal dependence structures associated with the unimodal and bimodal groups.

\subsection{Posterior of $Z_i,\pi_k$}

The posterior of $Z_i$ is discrete on $k=1,\dots,K$, with
\[
    P(Z_i=k\mid\dots)\propto
    \pi_k\,p(\bfpsi_i\mid Z_i=k,\dots).
\]
Here $p(\bfpsi_i\mid Z_i=k,\dots)$ is the normal density specified in the preceding subsection.

The mixing proportions have a stick-breaking prior $\GEM(\alpha)$, with $\alpha=1$ in practice. For the truncated $K$-component representation, let $V_k\sim\Beta(1,\alpha)$ independently for $k=1,\dots,K-1$ and set $V_K=1$. Then
\[
    \pi_k=V_k\prod_{b=1}^{k-1}(1-V_b),\qquad k=1,\dots,K,
\]
where the empty product for $k=1$ equals one. Writing $n_k$ for the number of observations in cluster $k$, the full conditional for $k=1,\dots,K-1$ is
\[
    V_k\mid Z,\alpha\sim
    \Beta\sbr{1+n_k,\,\alpha+\sum_{l=k+1}^K n_l}.
\]
In practice, draw independently
\begin{align*}
    G_{k,1}&\sim\text{Gamma}(1+n_k,1),\\
    G_{k,2}&\sim\text{Gamma}\sbr{\alpha+\sum_{l=k+1}^K n_l,1},
\end{align*}
using shape and rate parameterization, and compute $V_k=G_{k,1}/(G_{k,1}+G_{k,2})$.

\section{Competing methods used in the simulation}\label{supp_sec:sim_method}

\textbf{Figure~\ref{fig:ablation}.} We include the Dirichlet Multinomial Mixture (\texttt{DMM}) method \citep{holmes2012} in the simulation and real-data comparisons. DMM accommodates within-cluster heterogeneity through observation-specific probability vectors drawn from a common cluster-level Dirichlet distribution. Marginally, its count kernel is
\[
    X_i\mid Z_i=k\sim\text{Dirichlet-Multinomial}(N_i,\alpha_k),
\]
where $\alpha_k$ is the cluster-specific parameter vector.

IndTree models within-cluster heterogeneity through independent tree splitting variables, while CorTree additionally learns the covariance structure among those splitting variables. Thus the comparison of DMM, IndTree, and CorTree compares Dirichlet-based, independent-tree, and correlated-tree structures for within-cluster variation, as illustrated in Figure~\ref{fig:ablation}. Tables~\ref{tb:Sim_clus} and \ref{tab:RDA_ARI} report the corresponding clustering comparisons.

\begin{figure}[h]
    \centering    
    \begin{tikzpicture}[->, node distance=6cm, thick]
        \node (A) [rectangle, draw, minimum width=1cm, minimum height=0.8cm] {DMM};
        \node (B) [rectangle, draw, minimum width=1cm, minimum height=0.8cm, right of=A] {IndTree};
        \node (C) [rectangle, draw, minimum width=1cm, minimum height=0.8cm, right of=B] {CorTree};

        \draw (A) -- node[above, align=center] {+ tree-structured\\heterogeneity} (B);
        \draw (B) -- node[above, align=center] {+ flexible cluster-level\\covariance structure} (C);
\end{tikzpicture}
\caption{Incremental Differences of DMM, IndTree, and CorTree.}
\label{fig:ablation}
\end{figure}

\section{Additional Real Data Results} \label{supp_sec:RDA}

This section provides additional real data analysis results.

\begin{figure}[ht!]
    \centering
    \includegraphics[width=0.8\linewidth]{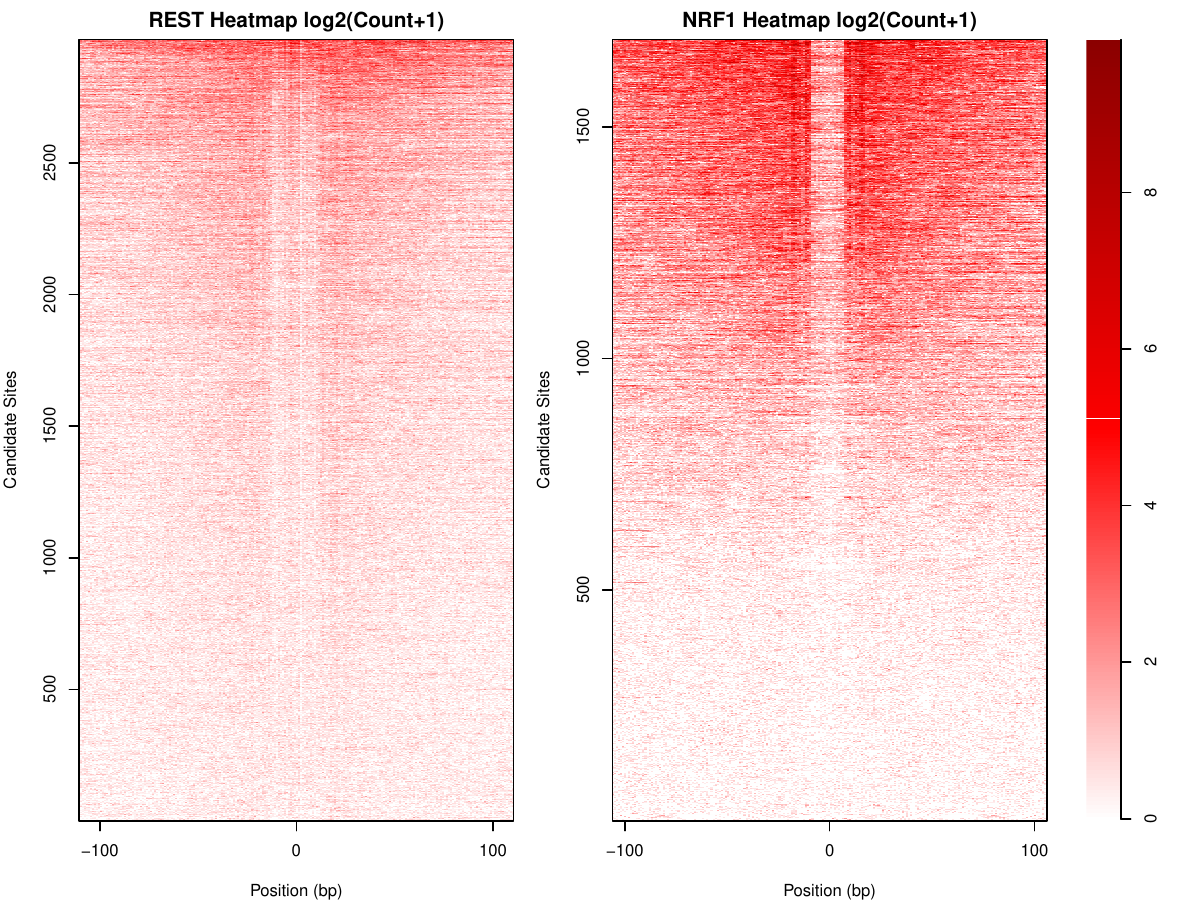}
    \caption{Heatmap of DNase seq data for REST and NRF1 (on chromosome 1 samples) in K562 (cell type). Each row represents one genomic region with motif match in the middle with 100 bp flanking regions on both sides of the motif, and each column represents the log-transformed ($\log_2(X+1)$) count of DNase-seq reads in the corresponding genomic location. The rows are ordered by the row sum in decreasing order.}
    \label{fig:heatmap}
\end{figure}

\textbf{Figure~\ref{fig:heatmap}} provides some illustration on how the real data $\bfX\in\bN_0^{n\times p}$ looks like.

\begin{figure}[ht]
\centering
\begin{subfigure}[t]{\textwidth}
  \centering
  \includegraphics[width=\linewidth]{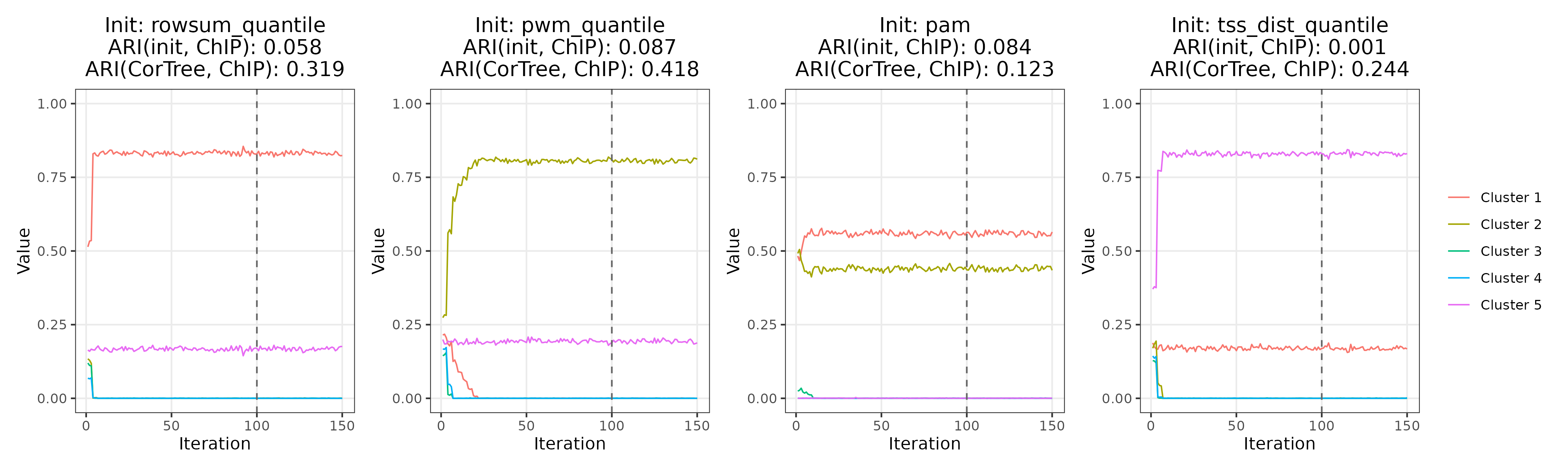}
  \caption{REST}
  \label{fig:REST_K_trace}
\end{subfigure}
\begin{subfigure}[t]{\textwidth}
  \centering
  \includegraphics[width=\linewidth]{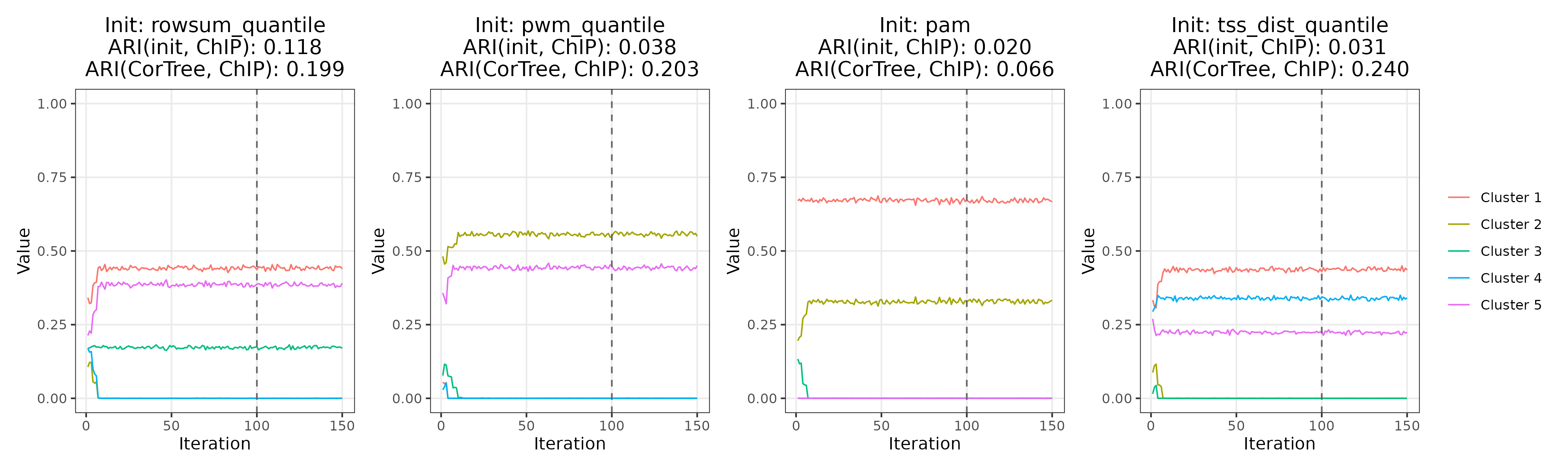}
  \caption{NRF1}
  \label{fig:NRF1_K_trace}
\end{subfigure}
\caption{REST and NRF1 traceplots for a total of 100 MCMC burn-in iterations and 50 additional iterations for the mixing weights $\pi$ (y-axis). Each subplot corresponds to a different initial value of the cluster membership assignment. The first column of subplots uses the quantiles of total count (row sums) as the initial value; the second column uses the quantiles of the PWM score as the initial value; the last column uses PAM clustering results as the initial value.}
\label{fig:mixing}
\end{figure}

\textbf{Table~\ref{tab:RDA_centipede}.} In the original CENTIPEDE paper \citep{pique2011accurate}, the predictors are the linear combination of $\text{tss\_dist}_i, \text{PWM\_score}_i$, and another variable for the evolutionary sequence conversation. In our main analysis, to keep a fair comparison, we only use $\text{PWM\_score}_i$ as the predictor since CorTree only uses PWM as the initial value, and CorTree clustering does not depend on any other covariates. In Table~\ref{tab:RDA_centipede}, we try different predictor combinations to align with \citep{pique2011accurate}. Compared to Table~\ref{tab:RDA_ARI} in the main manuscript, the REST result of CENTIPEDE is better than CorTree without PWM initialization, worse than CorTree using PWM initialization; the NRF1 result of CENTIPEDE is still worse than any CorTree results.

\begin{table}[ht]
\centering
\caption{ARI of CENTIPEDE using other predictors, with the binary ChIP label as reference truth.}
\label{tab:RDA_centipede}
\begin{tabular}{lrr}
\toprule
Predictors & $(1,\text{tss\_dist}_i)$ & $(1,\text{tss\_dist}_i, \text{PWM\_score}_i)$\\
\hline
REST & 0.302 & 0.336 \\
NRF1 & 0.089 & 0.089 \\
\bottomrule
\end{tabular}
\end{table}

\textbf{Table~\ref{tab:REST_nopwm}.} The PWM score itself is indeed good information about TF binding specificity, and using that could help improve the performance. Hence, we also want to test what happens if we do not use PWM score as a data preprocessing rule.
For reference, the range of PWM score in the raw NRF1 data is in $(13.89, 19.27)$, so the pwm-rule has no influence; the range of PWM score in the raw REST data is in $(10.18, 33.67)$, with a median at 12.11. In the supplementary, we rerun the REST analysis without the pwm-rule, where $\bfX_{\text{REST}}\in \bN_0^{11626\times 220}$. The results are shown in Table~\ref{tab:REST_nopwm}. In this larger data set, CorTree with all three initial values outperforms CENTIPEDE.

\begin{table}[ht]
\centering
\caption{ARI of REST experiment without using PWM as a preprocessing rule.}
\label{tab:REST_nopwm}
\begin{tabular}{lcccc}
\toprule
Method & K-means & PAM & IndTree & CENTIPEDE \\
ARI & 0.112 & 0.226 & 0.043 &  0.112 \\
\hline
Method&\shortstack[c]{CorTree \\ (init:\textit{rowsum})}& \shortstack[c]{CorTree \\ (init:\textit{pwm})}& \shortstack[c]{CorTree \\ (init:\textit{tss\_dist})}\\
ARI& 0.348 & 0.345 & 0.312\\
\bottomrule
\end{tabular}
\end{table}

\begin{table}[ht]
\centering
\caption{Elapsed time for REST and NRF1 analyses under different initialization methods. The last column is the computation time for DMM.}
\label{tab:RDA_elapsed_time}
\begin{tabular}{llrrr|r}
\hline
Dataset & Init & rowsum & pwm & tss-dist & DMM\\
\hline
REST & CorTree & 9.7 min & 9.6 min & 9.6 min & 2.3 min\\
     & IndTree & 9.5 min & 9.5 min & 9.5 min & -\\
\hline
NRF1 & CorTree & 1 hr 24.4 min & 1 hr 24.9 min & 1 hr 24.5 min & 1.7 min\\
     & IndTree & 1 hr 23.3 min & 1 hr 24.0 min & 1 hr 23.5 min & - \\
\hline
\end{tabular}
\end{table}

\begin{table}[ht]
\centering
\caption{Sensitivity analysis with varying hyperparameters.}
\label{tab:RDA_sensi}
\begin{tabular}{lclc}
\toprule
\multicolumn{2}{c}{REST}            & \multicolumn{2}{c}{NRF1}        \\
Hyperparameters   & CorTree ARI & Hyperparameters   & CorTree ARI \\
cutoff\_layer=4  & 0.418       & cutoff\_layer=3  & 0.163       \\
c\_sigma2\_vec=1 & 0.413       & c\_sigma2\_vec=1 & 0.203       \\
sigma\_mu2=1     & 0.417       & sigma\_mu2=1     & 0.226     \\
\bottomrule
\end{tabular}
\end{table}

\begin{figure}
    \centering
    REST\\
    \includegraphics[width=\linewidth]{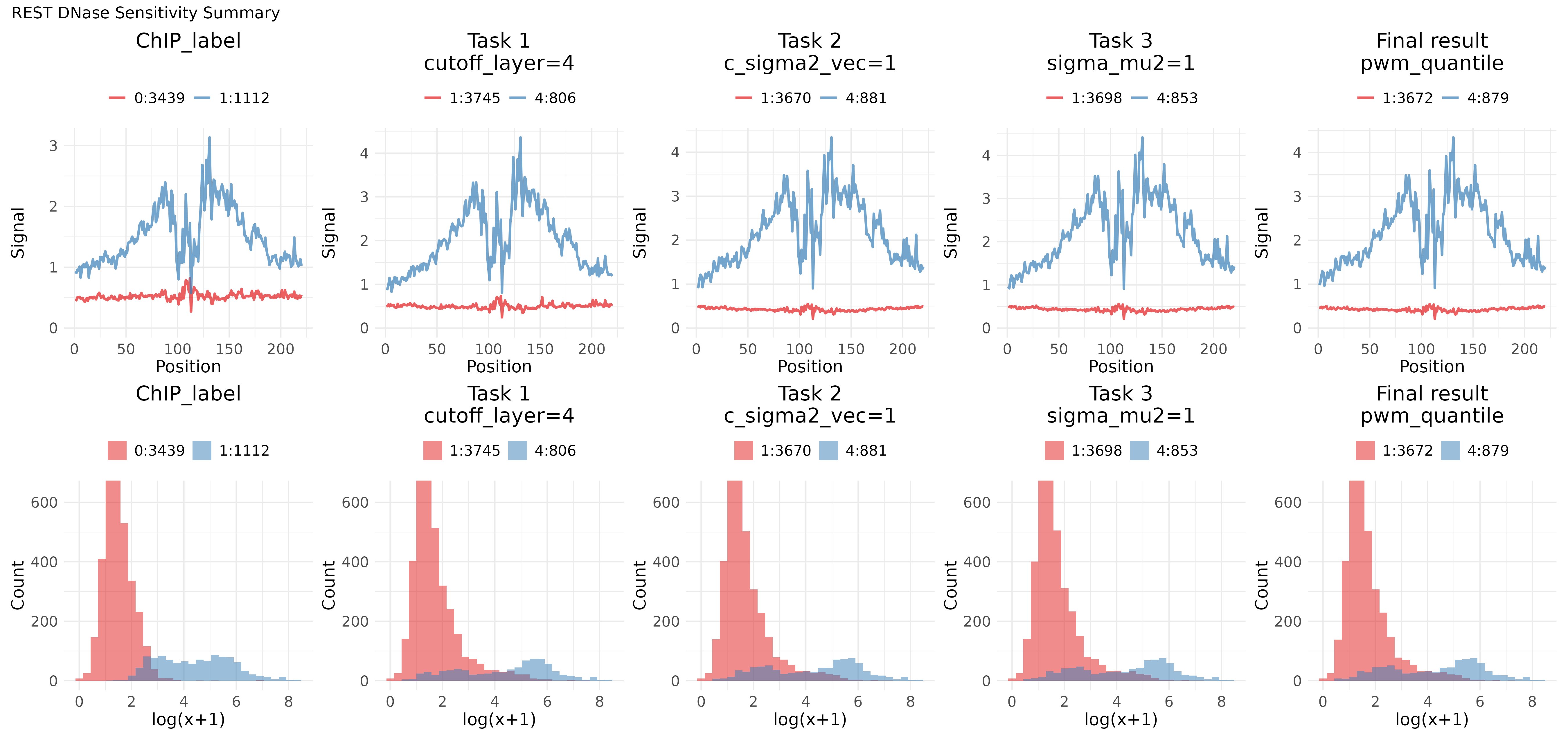}\\
    NRF1\\
    \includegraphics[width=\linewidth]{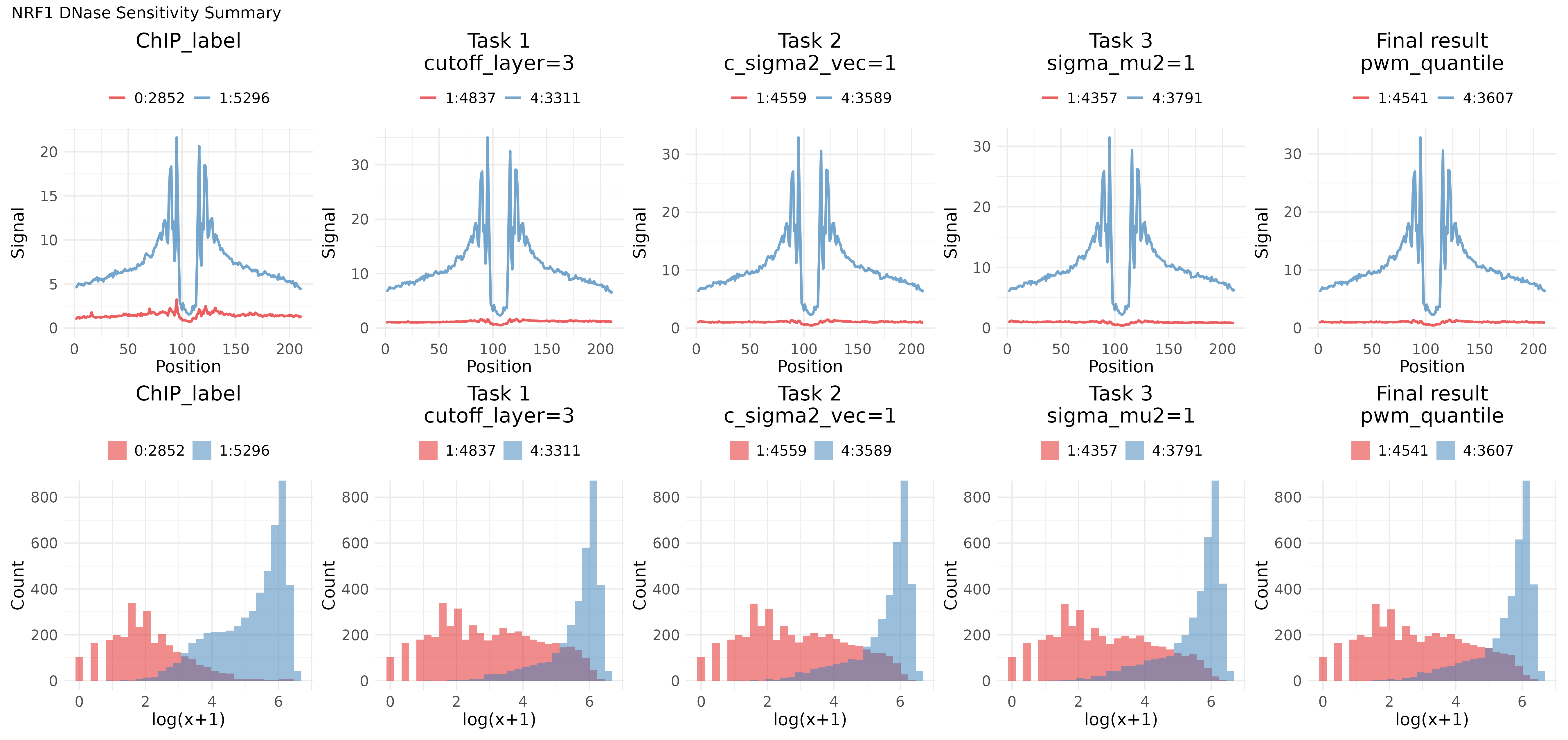}
    \caption{Sensitivity analysis results for REST and NRF1. Each column is one different hyperparameter compared with the main result setting in Table \ref{tab:RDA_sensi}. The final result and all sensitivity results are based on the PWM score as the initial value.}
    \label{fig:RDA_sensi}
\end{figure}

\subsection{Case study: American Gut Project}\label{supp_sec:AGP}
\begin{figure}
    \centering
    \includegraphics[width=\linewidth]{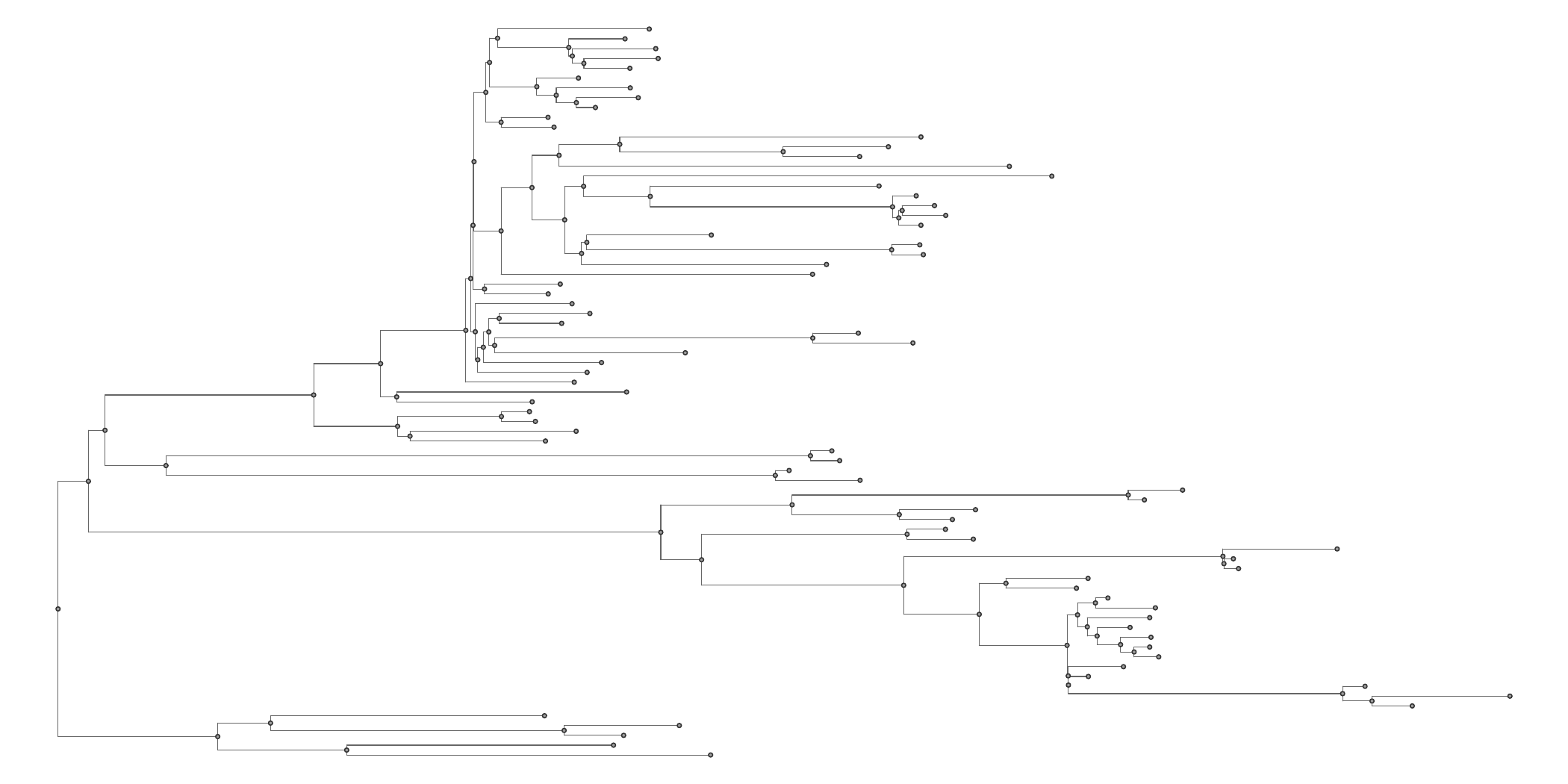}
    \caption{Phylogenetic tree used in AGP. Left to right: root node to leaf node.}
    \label{fig:phylogenetic_tree}
\end{figure}

In addition to the binary partition tree structure used to cluster the DNase-seq count data, to demonstrate the generalisability of the proposed CorTree framework, we have implemented a variation of CorTree, referred to as the \texttt{PhyloTree}. \texttt{PhyloTree} uses the phylogenetic tree structure instead of the binary dyadic tree partition. Figure~\ref{fig:phylogenetic_tree} is the visualization of the phylogenetic tree structure used in the AGP data. This tree organizes the microbes by evolutionary relatedness. Microbes that sit on nearby branches are more similar in their ancestry, while microbes that split apart earlier in the tree are less closely related. The observed data only contains the lead node. The count data is in the form of a count matrix $X$ of dimension $n\times K$, where $K$ is the number of taxa (leaf nodes).

\begin{figure}
    \centering
    \includegraphics[width=\linewidth]{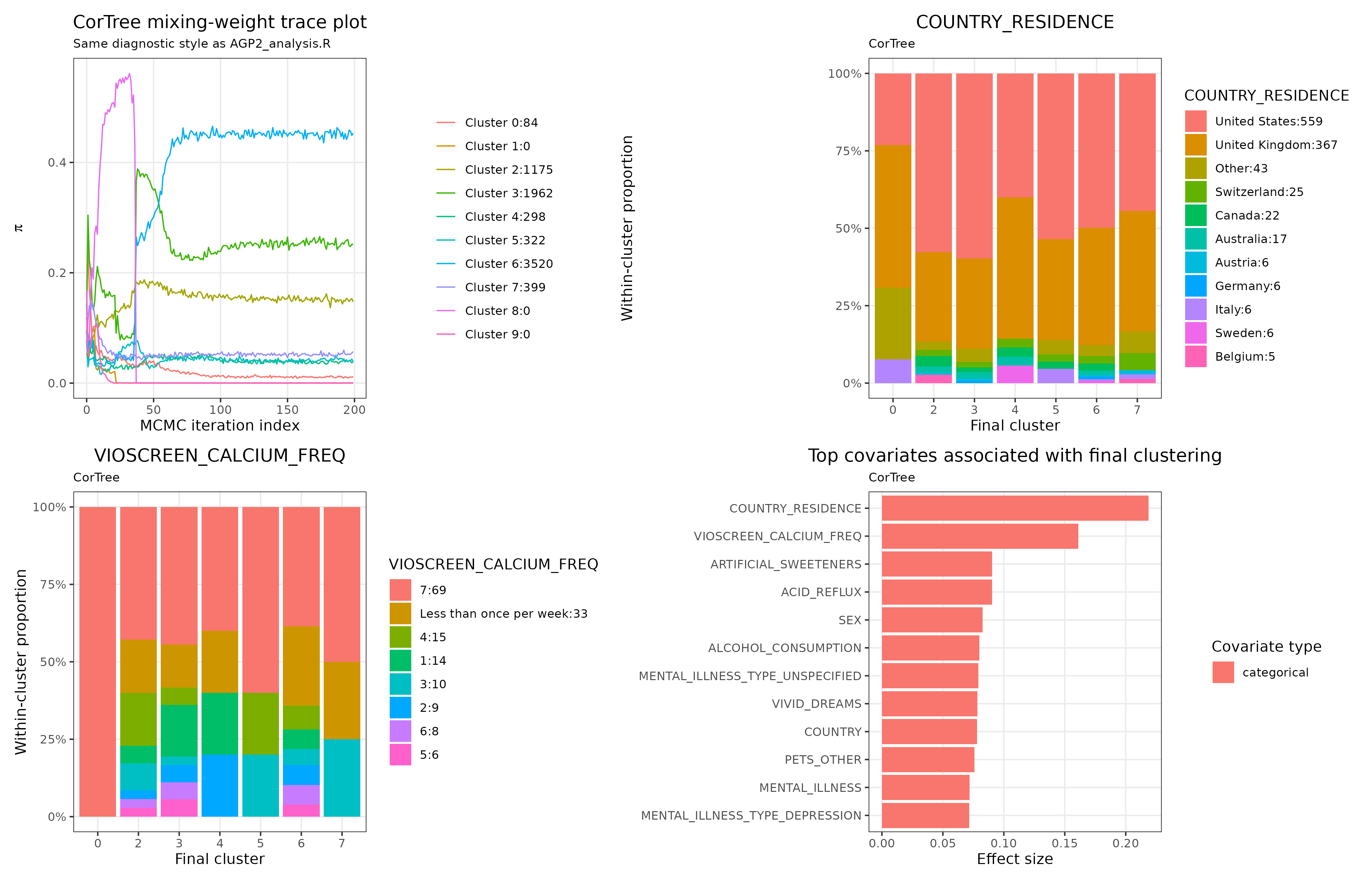}
    \caption{CorTree clustering result for the AGP data.}
    \label{fig:agp_cortree}
\end{figure}

\begin{figure}
    \centering
    \includegraphics[width=\linewidth]{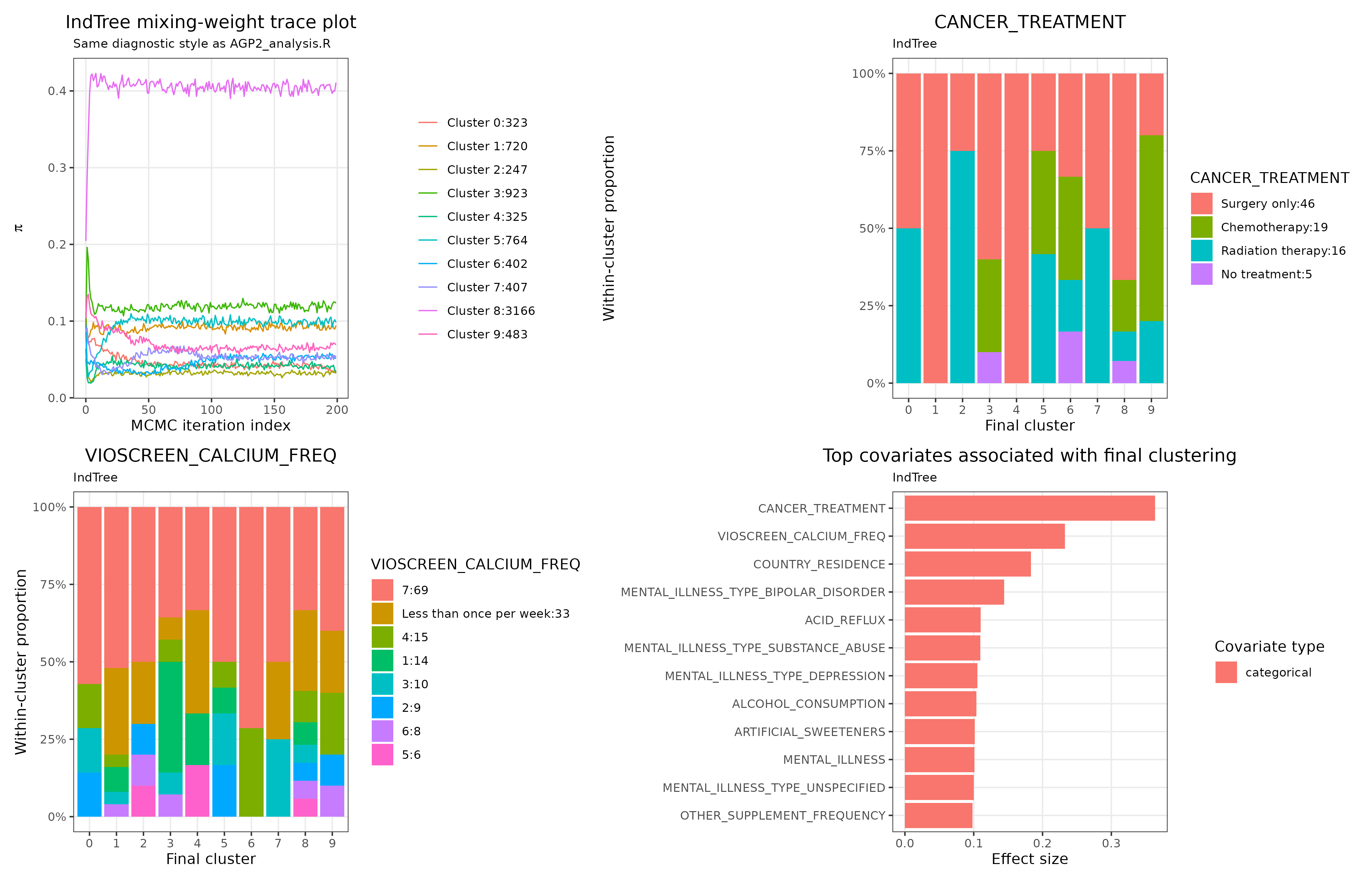}
    \caption{IndTree clustering result for the AGP data.}
    \label{fig:agp_indtree}
\end{figure}

\begin{figure}
    \centering
    \includegraphics[width=\linewidth]{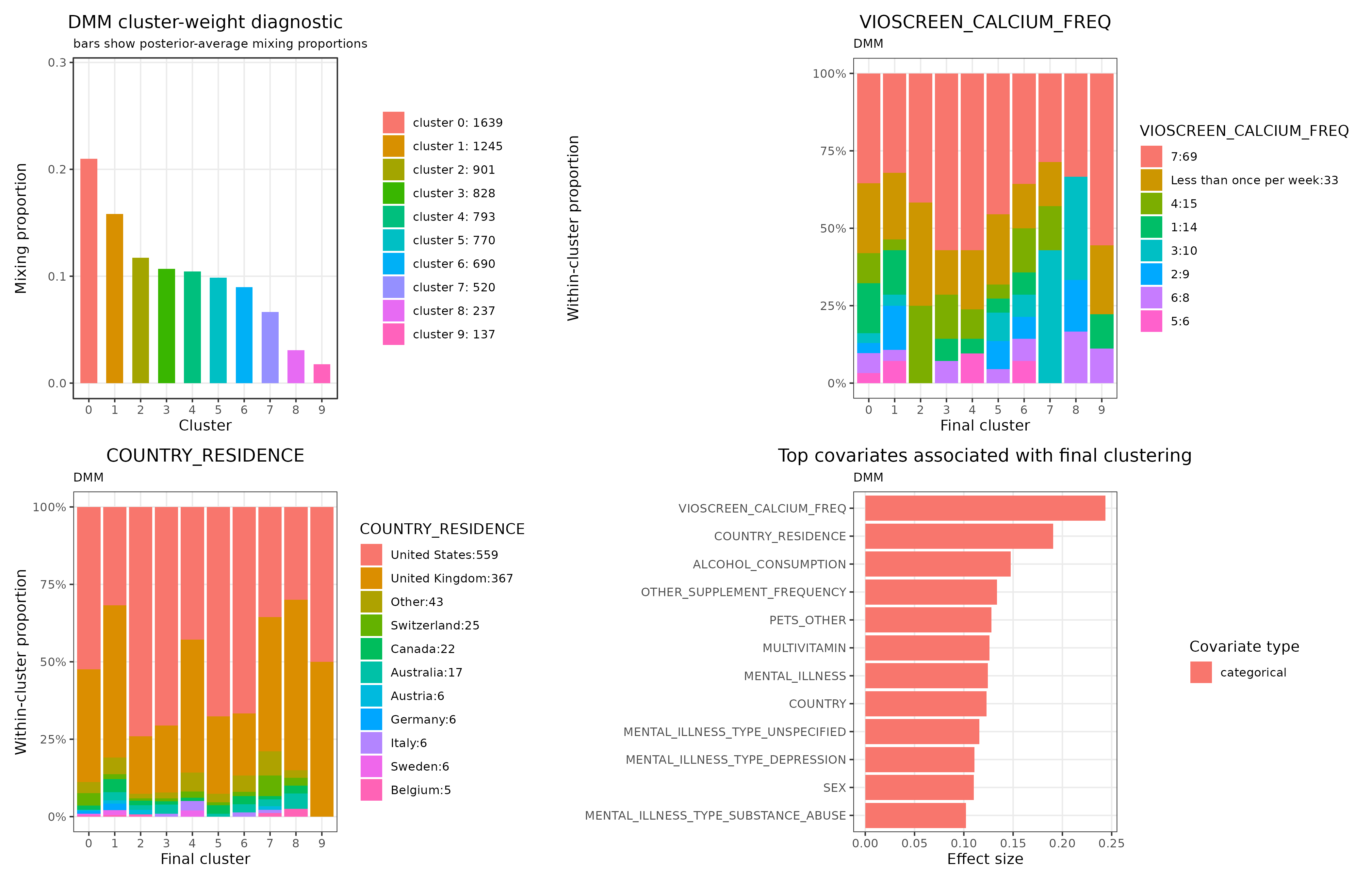}
    \caption{DMM clustering result for the AGP data.}
    \label{fig:agp_dmm}
\end{figure}

\begin{figure}
    \centering
    \includegraphics[width=\linewidth]{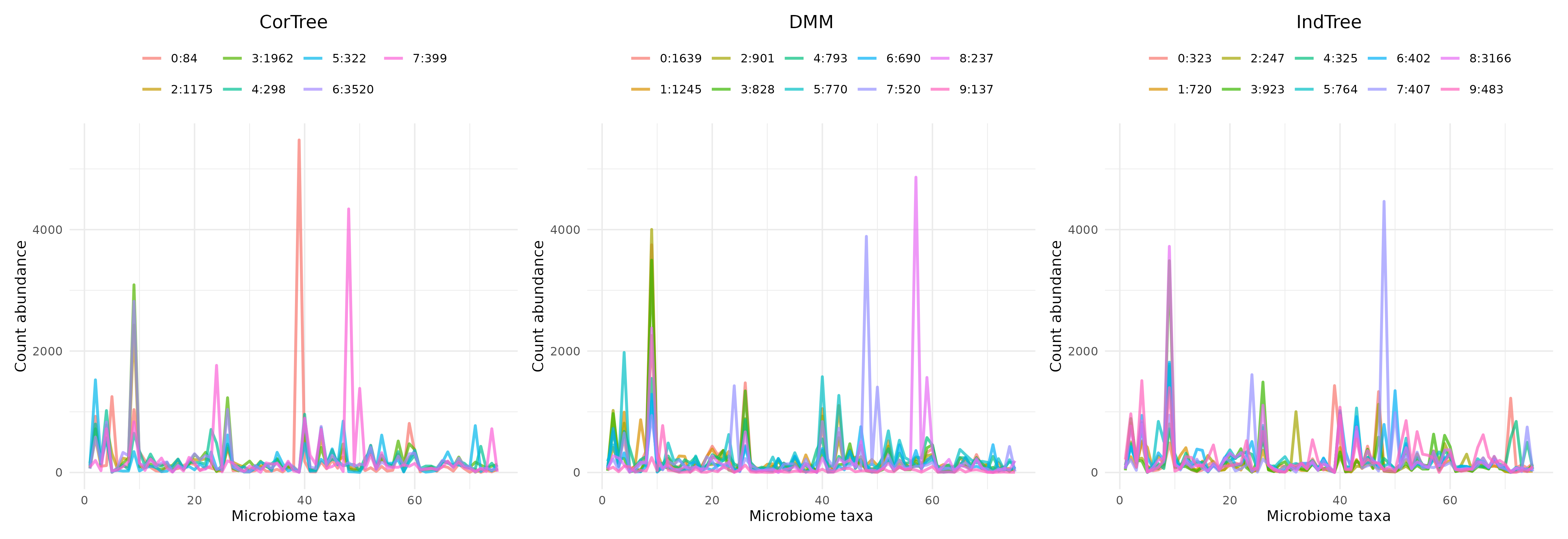}
    \caption{AGP cluster mean for each method}
    \label{fig:agp_mean}
\end{figure}

Figures~\ref{fig:agp_cortree} to \ref{fig:agp_dmm} show the clustering result of CorTree, IndTree, and DMM. Unlike the DNase-seq data, there's no reference truth for the microbiome clustering results. So we computed the strength of association (Cramer's V statistics) between each cluster result and the categorical covariates available in the AGP data, and ranked them by the effect size of the Cramér's V statistics (ranging from 0 to 1, with 0 indicating no association, and 1 indicating completely dependent). The most interesting result is in Figure~\ref{fig:agp_cortree}, where CorTree identifies a small subgroup 0 where most of the participants take calcium supplements 7 times a week. Frequent calcium intake can change the gut environment and further shape the microbiome composition profile.
Figure~\ref{fig:agp_mean} shows the cluster mean of each method across different microbiome taxa.

\clearpage
\codexnew{
\captionsetup{font={color=codexnewcolor},labelfont={color=codexnewcolor},hypcap=false}
\section{Numerical comparison with proper precision priors}\label{supp_sec:proper_priors}

\noindent\textbf{Prior specification and computation.}
We note that the original GHS prior \citep{li2019graphical} is improper: its flat diagonal factors integrate to infinity over the positive-definite matrices. In the flat-GHS implementation at GitHub commit \texttt{dc2c92d}, we stop updating the covariance when $\det(\Omega)\geq10^{200}$. After an update with $\det(\Omega)>10^{150}$, we add $10^{-15}I$ to the covariance before inversion. These rules limit extreme precision values but do not make the prior proper or establish posterior propriety. We retain the precision of an empty component because its full conditional under the flat prior is improper.

We compare flat GHS under the corrected sampler with three proper alternatives. We keep the horseshoe priors on off-diagonal entries and the prior on their common scale, and multiply the GHS prior by
\[
 w(\Omega)=
 \begin{cases}
 \prod_j\mathbf{1}\{0<\omega_{jj}<100\},&\text{bounded uniform},\\[1mm]
 |\Omega|^{a/2}\exp\{-\rho\operatorname{tr}(\Omega)\},&\text{weak/strong regularization}.
 \end{cases}
\]
We use $a=4$, $\rho=0.02$ for weak regularization and $a=N-N/K$, $\rho=0.005a$ for strong regularization, fixing both parameters throughout each analysis. The latter prior depends on sample size. With proper priors, we update empty-component precisions without the stopping or covariance adjustments above. Under the bounded uniform prior, we sample from the restricted conditional distributions rather than adjusting draws afterward.

\noindent\textbf{Simulation study.}
We simulate 100 datasets at each sample size ($N=200$, $400$, $600$), with two true clusters and 1,000 bins. We fit both tree models at depth 6 with three components, initialize allocations using total counts, and run 150 iterations, discarding the first 100. In CorTree, we correlate 31 splitting variables and update their covariance every five iterations; in IndTree, we update independent variances each iteration. We use the same data, initial allocations, and initial random-number state across CorTree priors. As in the original simulation code, we fix component counts within each profile, given its mixture weight.

\begin{center}
\begin{minipage}{\linewidth}
\centering\small\setstretch{1}
\captionof{table}{Mean ARI (standard deviation) over 100 simulated datasets per sample size. We use the corrected sampler for all tree-based results. Boldface marks the largest mean ARI in each row, including ties at the reported precision.}\label{tab:proper_prior_sim}
\setlength{\tabcolsep}{4pt}
\begin{tabular}{lccccc}
\toprule
Sample size & Flat GHS & \shortstack{Bounded\\uniform} & \shortstack{Weak\\regularization} & \shortstack{Strong\\regularization} & IndTree\\
\midrule
$N=200$ & \textbf{0.969} (0.057) & 0.967 (0.103) & \textbf{0.969} (0.068) & 0.891 (0.110) & 0.839 (0.075)\\
$N=400$ & 0.957 (0.152) & \textbf{0.971} (0.109) & 0.962 (0.127) & 0.885 (0.126) & 0.878 (0.060)\\
$N=600$ & 0.978 (0.104) & \textbf{0.981} (0.102) & 0.977 (0.103) & 0.938 (0.088) & 0.883 (0.050)\\
\bottomrule
\end{tabular}
\end{minipage}
\end{center}
We obtain similar mean ARIs with bounded uniform, weakly regularized, and flat GHS priors. Strong regularization reduces the mean ARI at every sample size.

\newpage
\noindent\textbf{DNase-seq analysis.}
We analyze 4,551 REST and 8,148 NRF1 sites. For both tree methods, we truncate the mixture at five components, use depth 9, and retain 50 of 150 MCMC iterations. For CorTree, we include parent-node depths 0--3 for REST and 0--4 for NRF1 in the correlated block and update its covariance every three iterations. For IndTree, we update the independent variances each iteration. We initialize allocations using total counts, PWM scores, or distance to the nearest transcription start site. 

\begin{center}
\begin{minipage}{\linewidth}
\centering\small\setstretch{1}
\captionof{table}{DNase-seq ARI. Both flat-GHS versions use the same prior; labels identify the precision-update sample size. The cluster-$n_k$ version also revises mixing-weight and other updates. Remaining methods use this sampler. Boldface marks row maxima.}\label{tab:proper_prior_dnase}
\setlength{\tabcolsep}{4pt}
\begin{tabular}{llrrrrrr}
\toprule
Dataset & \shortstack{Initial\\allocation} & \shortstack{Flat GHS\\(total $N$)} & \shortstack{Flat GHS\\(cluster $n_k$)} & \shortstack{Bounded\\uniform} & \shortstack{Weak\\regularization} & \shortstack{Strong\\regularization} & IndTree\\
\midrule
REST & Count & 0.319 & 0.0574 & 0.0576 & 0.0570 & \textbf{0.3349} & 0.0569\\
& PWM & \textbf{0.418} & 0.0724 & 0.0681 & 0.0702 & 0.1966 & 0.0706\\
& TSS & \textbf{0.244} & 0.0482 & 0.0494 & 0.0473 & 0.1067 & 0.0419\\
\midrule
NRF1 & Count & \textbf{0.199} & 0.1168 & 0.1161 & 0.1167 & 0.1968 & 0.1135\\
& PWM & \textbf{0.203} & 0.1684 & 0.1657 & 0.1679 & 0.1934 & 0.1586\\
& TSS & \textbf{0.240} & 0.1742 & 0.1846 & 0.1770 & 0.2162 & 0.1785\\
\bottomrule
\end{tabular}
\end{minipage}
\end{center}

Among the priors fitted with the corrected CorTree sampler, we obtain the highest DNase-seq ARI with strong regularization for each initial allocation. For REST, the ARI with total-count initialization is 0.3349, close to CENTIPEDE's 0.3364. GHS with stronger regularization improves DNase-seq clustering although it has lower simulation accuracy.

\noindent\textbf{Remark.}
We note that we obtain lower DNase-seq ARIs with flat GHS under the corrected sampler. Flat GHS with cluster $n_k$ is the corrected sampler based on the main-text GHS version (with total $N$): the corrected sampler uses component size $n_k$, rather than total $N$, with the within-component sum-of-squares matrix in the GHS update (see details in the GitHub repository). We have also corrected other sampling steps and changed the clustering summary. 

We provide the \href{https://github.com/yuliangxu/CorTree/blob/dc2c92d08779dcd8242760ebd285866867b22a23/docs/sampler-corrections.md}{numerical summaries} and \href{https://github.com/yuliangxu/CorTree/tree/dc2c92d08779dcd8242760ebd285866867b22a23/reproduction/ghs}{analysis scripts} at GitHub commit \texttt{dc2c92d}, including the random seeds, prior parameters, and diagnostic checks. The DNase-seq analyses require the processed data described there.\label{supp:proper_priors_end}
}

\end{document}